\documentclass{emulateapj}
\usepackage{apjfonts}

\newcommand{\rl}{$R_{\rm BLR}$--$L$}
\newcommand{\msigma}{$M_{\rm BH}-\sigma_{\star}$}
\newcommand{\mbh}{$M_{\rm BH}$}
\newcommand{\hst}{{\it HST}}
\newcommand{\sersic}{S\'{e}rsic}
\newcommand{\galfit}{{\sc Galfit}}

\shorttitle{Low-Luminosity End of the R--L Relationship}
\shortauthors{Bentz, et al.}

\received{2012 December 19}
\accepted{}

\begin{document}

\title{The Low-Luminosity End of the Radius--Luminosity Relationship
  for \\ Active Galactic Nuclei}

\author{ Misty~C.~Bentz\altaffilmark{1},
         Kelly~D.~Denney\altaffilmark{2,3},
         Catherine~J.~Grier\altaffilmark{4},
         Aaron~J.~Barth\altaffilmark{5},
         Bradley~M.~Peterson\altaffilmark{4,6},
	 Marianne~Vestergaard\altaffilmark{3,7}, 
         Vardha~N.~Bennert\altaffilmark{8},
	 Gabriela~Canalizo\altaffilmark{9},
         Gisella~De~Rosa\altaffilmark{4},
         Alexei~V.~Filippenko\altaffilmark{10},
         Elinor~L.~Gates\altaffilmark{11},
         Jenny~E.~Greene\altaffilmark{12},
         Weidong~Li\altaffilmark{10,13},
         Matthew~A.~Malkan\altaffilmark{14},
	 Richard~W.~Pogge\altaffilmark{4,6}, 
	 Daniel~Stern\altaffilmark{15},
	 Tommaso~Treu\altaffilmark{16}, and
	 Jong-Hak~Woo\altaffilmark{17}
}

\altaffiltext{1}{Department of Physics and Astronomy,
		 Georgia State University,
		 Atlanta, GA 30303, USA;
		 bentz@chara.gsu.edu}

\altaffiltext{2}{Marie Curie Fellow}

\altaffiltext{3}{Dark Cosmology Center,
                 Niels Bohr Institute,
                 Juliane Maries Vej 30,
                 2100 Copenhagen \O, Denmark}

\altaffiltext{4}{Department of Astronomy, 
		 The Ohio State University, 
		 140 West 18th Avenue, 
		 Columbus, OH 43210, USA}

\altaffiltext{5}{Department of Physics and Astronomy,
                 4129 Frederick Reines Hall,
                 University of California,
                 Irvine, CA 92697, USA}

\altaffiltext{6}{Center for Cosmology and AstroParticle Physics,
                 The Ohio State University,
                 191 West Woodruff Avenue,
                 Columbus, OH 43210, USA}

\altaffiltext{7}{Steward Observatory, 
                 University of Arizona, 
                 933 N Cherry Avenue, 
                 Tucson, AZ 85721, USA}

\altaffiltext{8}{Physics Department, 
                 California Polytechnic State University, 
                 San Luis Obispo, CA 93407, USA}

\altaffiltext{9}{Department of Physics and Astronomy, 
                 University of California,
                 Riverside, CA 92521, USA}

\altaffiltext{10}{Department of Astronomy, 
                 University of California,
                 Berkeley, CA 94720, USA}

\altaffiltext{11}{University of California Observatories/Lick Observatory, 
                 P.O. Box 85, 
                 Mount Hamilton, CA 95140, USA}

\altaffiltext{12}{Department of Astrophysical Sciences, 
                 Princeton University, 
                 Peyton Hall -– Ivy Lane, 
                 Princeton, NJ 08544, USA}

\altaffiltext{13}{Deceased 2011 December 12}

\altaffiltext{14}{Department of Physics and Astronomy, 
                  University of California, 
                  Los Angeles, CA 90095, USA}

\altaffiltext{15}{Jet Propulsion Laboratory, 
                  California Institute of Technology, 
		  4800 Oak Grove Drive, 
		  Pasadena, CA 91109, USA}

\altaffiltext{16}{Department of Physics, 
                  University of California, 
                  Santa Barbara, CA 93106, USA}

\altaffiltext{17}{Astronomy Program,
                  Department of Physics and Astronomy,
		  Seoul National University, 
		  Seoul, Republic of Korea}

\begin{abstract}
We present an updated and revised analysis of the relationship between
the H$\beta$ broad-line region (BLR) radius and the luminosity of the
active galactic nucleus (AGN).  Specifically, we have carried out
two-dimensional surface brightness decompositions of the host galaxies
of 9 new AGNs imaged with the {\it Hubble Space Telescope} Wide Field
Camera 3.  The surface brightness decompositions allow us to create
``AGN-free'' images of the galaxies, from which we measure the
starlight contribution to the optical luminosity measured through the
ground-based spectroscopic aperture.  We also incorporate 20 new
reverberation-mapping measurements of the H$\beta$ time lag, which is
assumed to yield the average H$\beta$ BLR radius.  The final sample
includes 41 AGNs covering four orders of magnitude in luminosity.  The
additions and updates incorporated here primarily affect the
low-luminosity end of the \rl\ relationship.  The best fit to the
relationship using a Bayesian analysis finds a slope of $\alpha =
0.533^{+0.035}_{-0.033}$, consistent with previous work and with
simple photoionization arguments.  Only two AGNs appear to be outliers
from the relationship, but both of them have monitoring light curves
that raise doubt regarding the accuracy of their reported time lags.
The scatter around the relationship is found to be $0.19
\pm\ 0.02$\,dex, but would be decreased to 0.13\,dex by the removal of
these two suspect measurements.  A large fraction of the remaining
scatter in the relationship is likely due to the inaccurate distances
to the AGN host galaxies.  Our results help support the possibility
that the \rl\ relationship could potentially be used to turn the BLRs
of AGNs into standardizable candles.  This would allow the
cosmological expansion of the Universe to be probed by a separate
population of objects, and over a larger range of redshifts.
\end{abstract}


\section{Introduction}

The ability to determine black hole masses is a crucial step toward
understanding the link between galaxies and their black holes, as well
as the details of the black hole environment.  To date, dynamical
methods have resulted in measurements of some 50 black hole masses.
However, these methods require that the gravitational influence of the
black hole on the stars or gas be spatially resolved, effectively
limiting the reach of current dynamical methods to galaxies no further
than $\sim 150$\,Mpc for even the most massive black holes (see
\citealt{gultekin09}).

Active galactic nuclei (AGNs), in contrast, are some of the most
luminous objects in the Universe, and are thus capable of providing us
with the leverage needed to probe the growth and evolution of black
holes at any significant cosmological distance.  However, AGNs are
also so rare that even the nearest are generally too distant for
current instruments to spatially resolve the radius of influence of
the black hole and provide a local calibration for their masses.
Instead, the most successful technique for measuring black hole masses
in AGNs is reverberation mapping \citep{blandford82}. Reverberation
mapping requires high-quality spectrophotometric monitoring of an AGN
over an extended period of time.  The line-emitting regions that give
rise to the characteristic AGN spectral signatures are photoionized by
the hot accretion disk around the black hole.  The continuum flux
(which arises from the accretion disk or very close to it) varies with
time, and these variations are echoed later by changes in the flux of
the broad emission lines.  The delay time between the continuum
variations and the broad-line variations can be measured by cross
correlation of the light curves and gives the light-travel time across
the broad-line region (BLR), or the radius of the BLR when multiplied
by the speed of light.  In effect, reverberation mapping substitutes
high temporal resolution for high spatial resolution, allowing us to
probe regions of gas that are only $\sim 0.01$\,pc in extent
(comparable to the inner region of the Oort cloud in our own Solar
System; \citealt{brown04}) in the centers of arbitrarily distant
galaxies.  Combining the BLR radius with the mean velocity of the BLR
gas, as measured from the Doppler broadening of the emission lines,
and assumptions or indirect estimates of the virial coefficient gives
a direct constraint on the black hole mass via the virial theorem.

The validity of reverberation masses has been upheld by several
independent lines of evidence.  A subset of objects in the
reverberation sample have measurements for several different emission
lines throughout the ultraviolet and optical portions of their
spectra, and the multiple emission lines show a virial behavior (e.g.,
\citealt{peterson99,peterson00a,kollatschny03,bentz10a}).  Two AGNs in
the current reverberation sample --- NGC\,3227 and NGC\,4151 --- are
sufficiently close that dynamical modeling has successfully determined
their black hole masses, and both the stellar dynamical masses
\citep{davies06,onken07} and gas dynamical masses \citep{hicks08}
agree with the reverberation-based masses within the uncertainties.
Furthermore, a fully general Bayesian modeling code has recently been
developed to analyze reverberation-mapping datasets and place limits
on the black hole mass and the BLR geometry and dynamics
\citep{pancoast11}.  When applied to the reverberation mapping data
for Arp\,151 \citep{brewer11} and Mrk\,50 \citep{pancoast12}, the
method recovers a black hole mass that is essentially the same value
as that determined from the reverberation method outlined above, for
standard assumptions of the virial coefficient
\citep{bentz09c,barth11}.

Reverberation mapping has yielded black hole masses for $\sim 50$ AGNs
thus far \citep{peterson04,bentz09c}. The BLR radius--luminosity
correlation ($R_{\rm BLR} \propto L^{\alpha}$) derived from this
reverberation sample is the basis for \emph{all} secondary techniques
used to estimate black hole masses in distant AGNs (e.g.,
\citealt{laor98,wandel99b,mclure02,vestergaard06}).  The power of the
\rl\ relationship comes from the simplicity of using it to quickly
estimate \mbh\ for large samples of objects, even at high redshift,
with only a single spectrum per object.

This simplicity has led to the \rl\ relationship being heavily
utilized in the literature.  A small sampling of studies that have
utilized the \rl\ relationship in the last few years includes
investigations of \mbh\ in the most distant quasars (e.g.,
\citealt{willott10,mortlock11}), black hole mass functions and
Eddington ratio distributions through cosmic history (e.g.,
\citealt{greene07a,vestergaard08,vestergaard09,kelly09,schulze10}),
cosmic evolution of black holes and their host galaxies (e.g.,
\citealt{woo08,merloni10,bennert10}), the nature of narrow-line
Seyfert 1 galaxies (e.g., \citealt{mathur12,papadakis10}), duty cycles
of quasars (e.g., \citealt{shankar09}), accretion properties of
various types of AGNs (e.g., \citealt{wang09,cao10,trump11}), studies
of relativistic jets and the jet-disk connection (e.g.,
\citealt{sambruna06,tavecchio07,ghisellini10}), studies of black holes
in dwarf and low-mass galaxies (e.g.,
\citealt{greene07b,dong07,thornton08}), and studies of optical
transients (e.g., \citealt{drake11}).

Because of the utility of the \rl\ relationship, much work has gone
into removing biases and noise from the reverberation database.
Previous determinations of the \rl\ relationship used luminosity
measurements from ground-based spectra and found the slope to be
$\alpha \approx 0.7$ \citep{kaspi00,kaspi05}.  To achieve the low
level of uncertainties in the flux calibration necessary for
reverberation mapping, a large (e.g., 5\arcsec~$\times$~7\arcsec)
spectroscopic aperture is typically employed.  Therefore, for all the
nearby objects with reverberation masses, a substantial fraction of
the observed luminosity is actually the result of the host-galaxy
starlight and not the AGN itself. The entire low-luminosity end of the
\rl\ relationship was overestimated in $L$, in effect, artificially
steepening the slope.

\citet{bentz06a} analyzed \hst\ ACS images of the nearest
reverberation-mapped AGNs and their host galaxies taken through the
F550M medium-band $V$ filter.  The flux contribution of starlight
through the ground-based spectroscopic monitoring aperture was
measured from each image, and the reverberation-mapping luminosities
were corrected accordingly.  The resultant \rl\ relationship was
found, as expected, to have a much flatter slope ($\alpha = 0.52 \pm
0.04$ compared to $\alpha = 0.67 \pm 0.05$), consistent with simple
photoionization expectations.  Consequently, all of the remaining
objects in the reverberation-mapped sample were imaged with ACS in a
similar manner with the intent of properly accounting for the
starlight in each object, even when that contribution was assumed to
be small.  While the slope of the relationship did not change much
($\alpha = 0.52 \pm 0.06$; \citealt{bentz09b}), the scatter in the
relationship was reduced from $\sim 40$\% to $\sim 35$\%.  The
scarcity of measurements anchoring the low-luminosity end, in
particular, then became apparent.

In the meantime, much effort has gone into replacing noisy and poorly
sampled reverberation datasets and increasing the overall range of BLR
radii probed.  The last several years in particular have seen a huge
amount of effort invested in reverberation-mapping experiments that
preferentially target AGNs with relatively low luminosities.  The Lick
AGN Monitoring Project (LAMP) campaign targeted low-luminosity AGNs to
more fully populate the low-luminosity end of the \rl\ relationship
and succeeded in measuring H$\beta$ BLR radii for 8 new AGNs
\citep{bentz09c}.  Multiple recent campaigns at MDM Observatory have
mainly focused on replacing poorly sampled or noisy reverberation
datasets with high signal-to-noise ratio (S/N), high temporal cadence
spectroscopy (\citealt{denney10,grier12b}) to allow better constraints
on the BLR structure and kinematics.

Given the number of updates, improvements, and additions to the
reverberation database, we undertook a full recalibration of the \rl\
relationship in an effort to provide a more accurately calibrated
relationship for the community to use when estimating black hole
masses in AGNs.  In particular, our new calibration is more accurate
at the low-luminosity end where $L_*$ and sub-$L_*$ galaxies tend to
reside.

We assume a standard $\Lambda$CDM cosmology of $H_0 =
72$\,km\,s$^{-1}$\,Mpc$^{-1}$, $\Omega_{\rm M}= 0.3$,
$\Omega_{\Lambda} = 0.7$ throughout this work.

\boldmath
\section{New H$\beta$ BLR Measurements}
\unboldmath

Recent reverberation-mapping campaigns have focused mainly on the
low-luminosity end of the \rl\ relationship and provide several new
H$\beta$ BLR measurements to the reverberation sample.  The
measurements come in three separate flavors: (1) replacement
measurements for targets of previous reverberation campaigns for which
the light curves were noisy or undersampled and led to poor or biased
determinations of the H$\beta$ BLR radius, (2) additional measurements
for targets of previous reverberation campaigns that already have
accurate H$\beta$ time-lag measurements, and (3) H$\beta$ BLR
measurements for new objects that have not been previously examined
with reverberation mapping.  We provide a brief summary of each of the
monitoring programs with the new results that we incorporate here.
The interested reader should refer to the original manuscripts
reporting the H$\beta$ BLR measurements for more details.

\paragraph{MDM 2005}
3C\,390.3 was the subject of a 2005 monitoring campaign at MDM
Observatory that resulted in an additional H$\beta$ radius and
luminosity measurement for this object \citep{dietrich12}.

\paragraph{MDM 2007}
\citet{denney10} describe the results of a 2007 monitoring program at
MDM and other observatories that had a goal of obtaining high-quality,
densely sampled light curves to search for velocity-resolved time lags
in the emission lines.  H$\beta$ BLR measurements were derived for six
AGNs through this program, three of which were replacements for
poor-quality datasets, and two of which were additional measurements
for AGNs with other reliable measurements.  The final object,
Mrk\,290, was new.  Mrk\,290 was also included in the 2008 LAMP sample
of AGNs (see below) but did not exhibit strong variations during that
campaign.  The inclusion of Mrk\,290 in the LAMP sample led to it
being included in the \hst\ Cycle 17 imaging campaign that we describe
below, from which we are able to derive the starlight correction to
the luminosity.

\paragraph{LAMP 2008}
The 2008 Lick AGN Monitoring Project (LAMP) targeted AGNs with
estimated black hole masses in the range $10^6$--$10^7$\,M$_{\odot}$.
Measurements of the H$\beta$ BLR radius were determined for eight new
objects, and an additional measurement of the H$\beta$ BLR radius in
the well-studied AGN NGC\,5548 was also determined \citep{bentz09c}.
NGC\,5548 was the only galaxy in the LAMP sample with the appropriate
\hst\ imaging to allow a host-galaxy starlight correction.  In the
next sections we detail the \hst\ Cycle 17 imaging program through
which we obtained the necessary images for the remainder of the LAMP
sample, the host-galaxy surface brightness modeling of those images,
and the derived starlight corrections to the ground-based
spectroscopic monochromatic luminosities at 5100\,\AA.

\paragraph{MDM 2010}
Additional measurements of H$\beta$ radii were determined for four
AGNs in the reverberation sample during a 2010--2011 campaign at MDM
and other observatories \citep{grier12b,peterson13}.  Two other AGNs,
Mrk\,6 and Mrk\,1501, were new targets and reliable H$\beta$ radii
were determined for them.  Unfortunately, there is no suitable \hst\
imaging from which to measure the starlight correction to the
spectroscopic luminosity.  We are therefore unable to include them in
this analysis of the \rl\ relationship.

The addition of nine new AGNs to the reverberation sample along with
11 replacement or additional datasets for previously monitored AGNs
allows us to revisit the calibration of the \rl\ relationship, and in
particular to examine the form of the relationship at the lower
luminosity end.  We include in this analysis all reverberation
datasets for which (a) there is a reliable time lag measured for the
H$\beta$ emission line, and (b) there is medium $V$-band (F547M or
F550M) \hst\ imaging available so the host-galaxy contribution to the
rest-frame 5100\,\AA\ flux can be determined and removed. Other
archival \hst\ images are available for some of the objects not
included here, but these images are not suitable for our analysis for
one of three reasons: (1) they were taken with a different filter and
therefore include emission lines from the galaxy and/or the
narrow-line region, both of which would have to be corrected, and they
would require assumptions about the unknown underlying stellar
populations in the galaxy, and/or (2) the exposures are too shallow to
accurately constrain the host-galaxy surface brightness profiles, or
(3) the images are heavily saturated in the nucleus, with strong
bleeding and a loss of information at the galaxy center.  We do not
include reverberation measurements of other Balmer lines in this
analysis because previous work (\citealt{kaspi00,bentz09c}) has shown
that there are differences in the mean time lags determined for
different Balmer lines, most likely caused by radiative-transfer
effects in the BLR clouds.  We are left with the sample of 41 AGNs
that are listed in Table~\ref{tab:sample}.

\section{\hst\ Observations and Data Reduction}

Nine AGNs with new H$\beta$ time lags were imaged during Cycle 17
(GO-11662, PI Bentz) with the Wide Field Camera 3 (WFC3) UVIS channel
through the F547M (Str{\"o}mgren $y$) filter ($\lambda_{c} =
5447$\,\AA\ and $\Delta \lambda = 650$\,\AA).  This imaging setup
allowed us to probe the continuum flux from the AGN and the host
galaxy while avoiding strong emission lines.  One orbit was dedicated
to each object, and each orbit was divided into two sets of exposures
separated by a dithering maneuver to improve the sampling of the WFC3
point-spread function (PSF) and facilitate in the rejection of cosmic
rays and detector artifacts (such as transient warm pixels).  To
maximize the dynamic range of the final images, each set of three
exposures was graduated in time, with exposure times of approximately
30\,s, 300\,s, and 690\,s.  We did not dither during an exposure
sequence to ensure that all three images were taken at the same
position.  Most of our targets were compact enough to fit on a single
chip of the UVIS channel, but for NGC\,6814 we employed a larger
dithering maneuver to ensure that there was no loss of information
because of the gap between the chips.  Details of the
\hst\ observations are given in Table~\ref{tab:hstobs}.

We were able to correct for saturation in the long exposures by making
use of the linear nature of charge-coupled devices (CCDs).  Saturated
pixels in the nucleus of each galaxy were identified in each image by
consulting the data quality frames from the \hst\ pipeline.  These
saturated pixels were clipped from the image and replaced by the same
pixels from a shallower, unsaturated exposure, but scaled up by the
exposure-time ratio.  Cosmic rays were cleaned with the Laplacian
cosmic ray identification package L.A.Cosmic \citep{vandokkum01}.  All
of the frames for a single object were then combined with the {\it
  multidrizzle} task to create a distortion-free image of each AGN
host galaxy.  The final combined, drizzled images are shown in
Figure~\ref{fig:apertures} with the ground-based spectroscopic
monitoring apertures overlaid.  It can easily be seen that the host
galaxy of each AGN contributes a significant amount of light within
the monitoring aperture.

\section{Galaxy Surface Brightness Decompositions}

An important component of calibrating the \rl\ relationship is
properly correcting the $L$ measurements for the contribution from
host-galaxy starlight.  The method we employed here is similar to that
described by \citet{bentz06a,bentz09b}, where the analysis of 32
galaxies in our sample is reported; it relies on using the software
program \galfit\ \citep{peng02} to model the surface brightness
profiles of the host-galaxy images.  \galfit\ is a nonlinear
least-squares two-dimensional image-fitting algorithm.  We used the
latest version of \galfit\ (Version 3) which allows for the modeling of
spiral arms, rings, and irregular shapes (see \citealt{peng10} for a
full description and various examples).

For the surface brightness decomposition of each of the 9 new \hst\
host-galaxy images in this work, we employed a tilted plane for the
background sky flux and a TinyTim \citep{krist93} model for the
unresolved AGN.  TinyTim models were generated for each specific AGN
by creating a model at the specific detector position of each of the
pointings and combining these models through {\it multidrizzle} in the
same way that the AGN images were combined.  Host-galaxy bulges,
disks, and bars were all fit with \citet{sersic68} profiles of the
basic form
\begin{equation}
\Sigma(r) = \Sigma_e \exp\left[-\kappa \left( \left(\frac{r}{r_e}\right)^{1/n} - 1\right)\right],
\end{equation}
where $\Sigma_e$ is the pixel surface brightness at the effective
radius $r_e$.  The \sersic\ index, $n$, has a value of 1 for an
exponential disk, 4 for a \citet{devaucouleurs48} profile, and 0.5 for
a Gaussian.  Bulge and bar components were modeled by allowing the
\sersic\ index to vary with no constraints while disks were modeled by
holding the \sersic\ index fixed at a value of 1.  Fits that resulted
in bulge \sersic\ indices outside of the range $\sim 0.1$--6 were
considered unphysical and therefore unacceptable.  In these cases, we
required multiple PSF models in the center of the galaxy, offset by
fractions of a pixel, to keep the \sersic\ index of the bulge from
running up the maximum value allowed by \galfit, $n=20$.  A high
\sersic\ index has a very peaky shape with strong wings, and can mimic
a PSF+sky model.  A runaway \sersic\ index in our galaxy fitting is
likely because of the well-known PSF mismatch that can occur between
TinyTim models and WFC3 images due to spacecraft ``breathing'' and/or
jitter, but could potentially be caused by any marginally resolved
nuclear flux from hot gas or star clusters.  We assume here that the
cause is PSF mismatch and ascribe all of the flux in these multiple
PSF models (which we assume are modeling a single physical component)
to the AGN itself, and we describe various tests of the validity of
this assumption below.

\galfit\ allows for surface brightness decompositions that can be as
simple or complicated as the user may wish.  The ultimate goal of the
surface brightness modeling in this project was to accurately remove
the AGN PSF, thereby creating an ``AGN-free'' image of each host
galaxy from which the starlight contribution could be measured.
\citet{bentz09b} assume uncertainties of 0.1\,mag in the measured
host-galaxy flux based in the range of acceptable models that could be
found to fit an image.  Here, we investigate the uncertainty in the
best-fit models by carrying out two independent sets of surface
brightness decompositions for each of the host-galaxy images in this
study.

The first set of models, which we will refer to as the ``optimal''
models, include multiple surface brightness components and make use of
power-law rotation to model spiral arms, Fourier modes to account for
``bending'' of the ellipse modeling the light distribution and other
asymmetric flux distributions, and truncation functions to allow for
the modeling of rings.  These models, which are shown in the bottom
panels of Figures~\ref{fig:m142}$-$\ref{fig:n6814}, are the best
representations for the actual two-dimensional surface brightness
distributions of the host galaxies.

The second set of models, which we refer to as ``simple'' models, do
not make use of power-law rotations, Fourier modes, or truncation
functions.  The simple models typically require a factor of 3 fewer
free parameters than the optimal models and they are computationally much
faster to run and to converge, but they less accurately represent the
two-dimensional surface brightness profiles of the AGN host galaxies,
as can be seen in the top panels of
Figures~\ref{fig:m142}$-$\ref{fig:n6814}.  Conversely, the optimal models
do a good job of reproducing the relative flux in each pixel in the
images, but the importance of this, other than being aesthetically
pleasing, is not clear and the physical interpretation of each model
component is not straightforward to determine.

Tables~\ref{tab:m142}$-$\ref{tab:n6814} give the parameters determined
for the ``optimal'' and ``simple'' fits for each of the 9 galaxies fit
here.  The formats of the tables are as follows: Column (1) gives the
note for the type of fit described (``optimal'' or ``simple''); column
(2) gives the component number of the fit, generally in order of
increasing angular size and increasing angular offset from the center
of the galaxy; and column (3) gives the description for the type of model
component (or components in the case of a PSF model and tilted plane
sky model).  The remaining columns describe the various parameters of
each model, with Column (11) listing any notes relevant to the models.
We give a brief description of the remaining columns below, but the
interested reader is referred to \citet{peng10} for further details of
the models and their parameters employed by \galfit.

For the PSF models, columns (4) and (5) are the angular offsets in
arcseconds from the center of the galaxy (defined as the location of
the AGN PSF) in the $x$ and $y$ directions, respectively.  Column (6)
is the integrated magnitude of the PSF model.  For the sky models,
column (8) gives the average value of the sky background level in
counts at the geometric center of the image, and columns (9) and (10)
give the flux gradients in the $x$ and $y$ directions respectively.

\sersic\ models are listed with columns (4) and (5) as the angular
offsets in arcseconds from the center of the galaxy in the $x$ and $y$
directions, respectively.  Column (6) gives the integrated magnitude
of the \sersic\ component and column (7) lists the effective radius in
arcseconds.  Column (8) gives the \sersic\ index, which was held fixed
at a value of 1.0 for exponential-disk components.  Columns (9) and
(10) are the axis ratio and the position angle of the major axis in
the image.  Note that images were fit at the orientation obtained
during the observation, and the position angles listed would need to
be corrected for the roll angle of the spacecraft to determine their
orientation relative to north.

For the ``optimal'' fits, the \sersic\ models were modified by power-law
rotation, Fourier modes, and/or radial truncation functions.  In the
case of the truncation functions, where portions of the underlying
model are removed, the \sersic\ profile is listed as ``sersic3'' and
we report the surface brightness at the break radius ($\Sigma_{\rm
b}$) and the break radius ($r_{\rm b}$) itself in columns (6) and (7),
rather than the integrated magnitude and effective radius.

Power-law rotations of a \sersic\ profile are denoted by ``power'' in
column (3).  Columns (5) and (6) list the inner and outer radii of
rotation in arcseconds.  Column (7) gives the rotation angle between
the inner and outer radii and column (8) is the power-law slope,
denoted as $\alpha$.  Columns (9) and (10) are the line-of-sight
inclination angle of the disk, $\theta_{\rm incl}$ (with $\theta_{\rm
incl}=0$ being equivalent to face-on), and the position angle of the
rotation in the plane of the sky, $\theta_{\rm sky}$.

Fourier-mode modifications to \sersic\ profiles are denoted by
``Fourier'' in column (3).  Beginning with columns (5) and (6), and
continuing through column (10), are the modes (e.g., $m=1$ or $m=3$)
and their amplitudes $a_{\rm m}$ and phase angles ($\phi$), where the
phase angle is the relative angle between the Fourier mode and the
position angle of the major axis of the parent \sersic\ profile.  To
avoid degeneracy with the axis-ratio parameter for the
\sersic\ profiles, we did not make use of the $m=2$ Fourier mode.

Truncation functions were generally used to model rings in the
galaxies and are denoted as ``radial'' in column (3).  Both inner and
outer truncations were used, with each denoted appropriately.  Columns
(4) and (5) give the angular offsets of the center of the truncation
function from the center of the parent \sersic\ profile in the $x$ and
$y$ directions, respectively.  Column (7) gives the break radius of
the truncation function, defined to be the radius at which the
truncation function has a value of 99\% of the flux of the untruncated
\sersic\ model at that same radius.  Column (8) lists the softening
length, $\Delta r_{\rm soft}$, where $r_{\rm break} \pm \Delta r_{\rm
  soft}$ ($+$ for outer truncations, $-$ for inner truncations) gives
the radius at which the flux drops to 1\% of the untruncated
\sersic\ model flux.  Columns (9) and (10) give the axis ratio and
position angle of the truncation function.

Finally, the last row of each fit gives the figures of merit for that
particular surface brightness decomposition: $\chi^2$; the number of
degrees of freedom, $N_{\rm dof}$; the number of free parameters in the
models, $N_{\rm free}$; and the reduced $\chi^2$, $\chi_{\nu}^2$.

In addition, to test the suitability of our choice of using multiple
TinyTim PSFs offset by fractions of a pixel to better model the AGN
PSF in several objects, we carried out surface brightness
decompositions using a variety of different PSF models.  These
included a very high S/N WFC3 image of the white dwarf EGGR\,102, a
bright field star from the image of Mrk\,290 (one of our targets), a
fainter star in the field around EGGR\,102, and \citet{moffat69} fits
to each of these stars.  We also investigated the effect of convolving
the image and the PSF model with a narrow Gaussian to ensure Nyquist
sampling (e.g., \citealt{kim08}).  In each of these tests, we only
allowed a single component to model the AGN PSF and we compare the
results of the test to the results obtained using the fitting
procedures described above and tabulated in
Tables~\ref{tab:m142}$-$\ref{tab:n6814}.  When the image of a star was
used as the PSF model, the difference in central host-galaxy flux
measured from an ``AGN-free'' image was only $\sim 1$\%, and it was
only slightly higher ($\sim 2$\%) when a Moffat fit to a star image
was used as the PSF model. Broadening the image and the PSF model
caused the largest difference in central host-galaxy flux, about 7\%,
so while this approach has been found to work in the past for other
\hst\ cameras and in other situations, it was the least successful
alternative in this case.  As expected, the fit residuals at the
center of the galaxy are the smallest when we allow multiple TinyTim
models to account for the central AGN PSF.  Furthermore, TinyTim has
the advantage of producing PSF models with infinite S/N, therefore
avoiding the problem of introducing additional noise into the
``AGN-free'' images from which we determine the host-galaxy starlight
contribution.  We find the same results when we also fit the image of
the bright star in the field of Mrk\,290 with multiple TinyTim PSF
models --- the residuals in the fit are decreased without adding extra
noise.

Below, we provide some notes on each of the individual galaxies
modeled in this work.

\paragraph{Mrk\,142}
Mrk\,142 is a late-type spiral galaxy at intermediate inclination.  It
does not appear to have a bulge, but does seem to have a bar that
manifests itself as a compact structure with a low \sersic\ index
($n<1$) and elongated shape.  At the highest redshift of any of the
galaxies modeled here ($z=0.045$), the bulge may be too compact to
disentangle from the very bright unresolved AGN.  Its morphological
classification according to these images is SBcd-SBd.  The parameters
for the optimal and simple surface brightness decompositions of Mrk\,142
are tabulated in Table~\ref{tab:m142}, and the models and residuals
are displayed in Figure~\ref{fig:m142}.

\paragraph{SBS\,1116+583A}
SBS\,1116+583A is a relatively face-on barred spiral galaxy with an
exponential bulge, approximately SBb in type.  The best-fit parameters
for its surface brightness decompositions, which are given in
Table~\ref{tab:s1116}, include a lens (uniform disk,
\citealt{devaucouleurs59}) that is more extended than the bulge,
nearly circular, and has a very low \sersic\ index ($n \approx
0.3$--0.4) in both the ``optimal'' and ``simple'' models.  The models and
residuals are displayed in Figure~\ref{fig:s1116}.

\paragraph{Arp\,151}

Arp\,151 (Mrk\,40) is an early-type spiral galaxy (S0-Sa) with a hint
of remaining spiral structure and a long tidal tail stretching
north-northwest from a recent encounter with a small companion galaxy
at a projected angular distance of $\sim 19$\arcsec.  The messy
morphology of Arp\,151 and its companion required multiple surface
brightness components for an accurate fit, and we do not attempt an
interpretation of their physical meaning here.  The best-fit
parameters for its surface brightness decompositions are listed in
Table~\ref{tab:arp151}, and the models and residuals are displayed in
Figure~\ref{fig:arp151}.

\paragraph{Mrk\,1310}

Mrk\,1310 is a ringed spiral galaxy, approximately Sb in type, with an
apparently large number of bright globular clusters. There is also a
faint galaxy directly south of Mrk\,1310 that appears as an arc.  The
location of the galaxy along the line of sight to Mrk\,1310 is
unknown, but the distorted shape of this faint galaxy may mean that it
is being tidally disrupted by Mrk\,1310, or it may simply be a chance
superposition..  The large angular separation and orientation of
elongation rule out the possibility of gravitational lensing.
Additional color information, at the minimum, will be necessary to
determine where this small galaxy exists along our line of sight to
Mrk\,1310.  The best-fit parameters for the surface brightness
decompositions of Mrk\,1310 are given in Table~\ref{tab:m1310}, and
the models and residuals are displayed in Figure~\ref{fig:m1310}.

\paragraph{Mrk\,202}
Mrk\,202 is a compact face-on spiral galaxy, approximately Sb in type,
with a bright star-forming ring.  The best-fit parameters for its
surface brightness decompositions are listed in
Table~\ref{tab:m202}, and the models and residuals are displayed in
Figure~\ref{fig:m202}.

\paragraph{NGC\,4253}
NGC\,4253 (Mrk\,766) is a barred spiral galaxy of type SBc with a
distinct nuclear spiral and a faint outer ring.  It is also classified
as a narrow-line Seyfert~1 because of its relatively narrow broad
emission lines.  The best-fit parameters for its surface brightness
decompositions are given in Table~\ref{tab:n4253}, and the models
and residuals are displayed in Figure~\ref{fig:n4253}.

\paragraph{NGC\,4748}
NGC\,4748 is a barred spiral galaxy with a nuclear starbursting ring
and is currently undergoing an interaction with another, slightly
smaller, spiral galaxy.  The best-fit parameters for its surface
brightness decompositions are given in Table~\ref{tab:n4748}, and
the models and residuals are displayed in Figure~\ref{fig:n4748}.

\paragraph{Mrk\,290}
Mrk\,290 is an early-type spiral galaxy (Sa-Sab) at a relatively low
inclination to our line of sight.  The bright source to the southeast
appears to be a star in our own galaxy.  The best-fit parameters for
its surface brightness decompositions are listed in
Table~\ref{tab:m290}, and the models and residuals are displayed in
Figure~\ref{fig:m290}.

\paragraph{NGC\,6814}
NGC\,6814 is a beautiful, face-on, moderately barred spiral galaxy at a
fairly low redshift of 0.0052 ($D_L \approx 20$\,Mpc), making it
one of the nearest broad-lined AGNs in the local universe and in our
sample.  The best-fit parameters for its surface brightness
decompositions are given in Table~\ref{tab:n6814}, and the models
and residuals are displayed in Figure~\ref{fig:n6814}.

\section{AGN Fluxes}

\subsection{Starlight Measurements and AGN Flux Recovery}
The host-galaxy starlight contribution to the 5100\,\AA\ spectroscopic
flux was determined by first measuring the yield of electrons within a
rectangular aperture, with dimensions and orientation matching that of
the ground-based monitoring campaign, centered on the nucleus of the
galaxy in the PSF- and sky-subtracted {\it Hubble Space Telescope
  (HST)} image.  The exposure time and inverse sensitivity for each
image ({\it HST} keyword {\it photflam}, having units of
ergs\,cm$^{-2}$\,\AA$^{-1}$\,electron$^{-1}$) were utilized to recover
the incident photon flux from the yield of electrons.  All {\it
  photflam} values were taken from the most recent recalibration of
the appropriate dataset through the \hst\ pipeline as of 2012 June 12.
For the ACS images, the {\it photflam} values are somewhat different
from those previously used by \citet{bentz09b} because they have been
updated to account for the loss of sensitivity of the High Resolution
Channel over time \citep{bohlin12}.

Once the host-galaxy flux through the \hst\ system had been
determined, a small color correction was necessary to account for the
difference between rest-frame 5100\,\AA\ and the pivot
wavelength\footnote{A measure of the effective wavelength of a filter
  that is independent of the source spectral energy distribution
  \citep{tokunaga05}.} of the filter.  To determine the color
correction, a bulge template spectrum \citep{kinney96} was redshifted
and reddened by the appropriate amounts to approximate the central
host galaxy of each AGN.  Only Galactic extinction was included in the
reddening, and the values used are slightly different from previous
values employed by \citet{bentz09b} because they are based on the
\citet{schlafly11} recalibration of the \citet{schlegel98} dust map.
The Galactic extinction values are smaller by a few hundredths of a
magnitude for all of our sample (median difference of $-0.024$\,mag)
except for 3C\,120 where the new extinction value is 0.2\,mag smaller
than before.  The ratio of the 5100\,\AA\ flux to the flux through the
\hst\ filter (which we associate with the pivot wavelength) was
estimated from the redshifted and reddened spectrum using {\it
  synphot}, and is listed in Table~\ref{tab:flux}.  The final
derived host-galaxy flux contributions to the ground-based
spectroscopic continuum flux are given in Table~\ref{tab:flux}.  These
values were subtracted from the absolute calibrations of the mean
continuum fluxes during the monitoring campaigns to recover the mean
AGN fluxes at rest-frame 5100\,\AA, from which the AGN luminosities
were determined.  We discuss the effects of ground-based seeing and
modeling uncertainties, among others, below.

\subsection{Uncertainties}
The uncertainty in the recovered AGN flux is a combination of the mean
measurement uncertainty in the continuum flux from the reverberation
campaign and the uncertainty in the host-galaxy contribution to the
continuum flux.  The former is a small component, ranging from $1-5$\%
for the measurements included here.  This is due to the requirement
for reverberation-mapping campaigns to achieve a high S/N
per pixel in the continuum flux of each individual spectrum acquired
throughout the campaign (typically $S/N \approx 30$--100) in order to
measure the few-percent variations that evidence the reverberation
signal.  The latter contribution to the AGN flux uncertainty was
determined by adding in quadrature the uncertainty in the starlight
flux from the modeling and the uncertainty in the starlight flux from
ground-based seeing effects that would be in place during a
reverberation-mapping campaign.  

\paragraph{Modeling Uncertainties} 
The uncertainty from surface brightness modeling was determined by
comparing the starlight measurement derived for each object from the
simple fit and the best fit detailed in the previous section.  The
addition of Fourier modes and power-law rotation, in general, changed
the starlight measurement by 0.04\% (Mrk\,202) to 8\% (Mrk\,766), with
a median difference of 3\%.  Because such comparisons are
time-intensive and computationally demanding, we have not carried them
out for all 41 galaxies in the sample.  Instead, we adopt a
conservative estimate of 5\% uncertainty for the host-galaxy
contribution for all compact galaxies, where the field of view of the
\hst\ image contains a large fraction of pixels that consist of empty
sky (e.g., the Markarian objects and the PG quasars).  For extended
galaxies that fill the field of view of the \hst\ camera with which
they were observed (e.g., the NGC galaxies with the ACS HRC), we adopt
a 10\% uncertainty in the host-galaxy contribution due to the greater
uncertainty in the determination of the background sky level during
the modeling process.

\paragraph{Seeing Effects}
Optical reverberation-mapping campaigns are generally carried out from
the ground and thus have to contend with variable seeing from night to
night throughout a campaign.  The effects of slit losses and variable
seeing on the measured AGN flux are minimized by using a wide
spectroscopic slit (4\arcsec--5\arcsec) and by carrying out an
internal calibration of all the spectra obtained for an object by
utilizing the non-variable [\ion{O}{3}] $\lambda \lambda$ 4959,~5007
doublet.  Nevertheless, seeing redistributes the galaxy flux as well
and can cause the starlight measurements from diffraction-limited
\hst\ images to differ from the contribution obtained through the
ground-based setup under typical seeing conditions. Reverberation
campaigns generally scale the final spectra to the [\ion{O}{3}] flux
measured on photometric nights, with a typical seeing of $\sim
1$\arcsec.  To investigate the effect of seeing on the derived
host-galaxy flux, we took the ``AGN-free'' images of NGC\,5548 (an
extended galaxy) and SBS\,1116+583A (a compact galaxy) and created a
simulated ground-based image of each by smearing with a 1\arcsec\ FWHM
Gaussian.  The starlight measurements were then made in the same way
from the simulated ground-based images as they were from the
diffraction-limited images.  The difference in measured starlight flux
was negligible for NGC\,5548, only 2\%, but was 8\% for
SBS\,1116+583A.  Based on these results, we adopt an average 5\%
uncertainty for the host-galaxy contribution for each object in the
sample due to ground-based seeing effects.

\paragraph{Background Determination}
Finally, we have also considered the effect of background subtraction
during the spectral reductions on the host-galaxy flux.  For the
extended galaxy NGC\,5548, we measured the host-galaxy flux in the
``background'' regions on either side of the extraction region.  The
average of these background regions was treated as ``sky'' flux and
subtracted from the flux within the extraction region.  The difference
in host-galaxy flux was found to be only 2\% even though NGC\,5548 is
a bright extended galaxy.  For the more compact galaxies in our
sample, the effect would be even less.  Therefore, we consider the
effect of background-subtraction regions during spectral reductions to
be negligible.

\section{Distances and AGN Luminosities}

By far, the largest contribution to the uncertainty of the AGN
luminosities is from the uncertain distance to each AGN.  Only five of
the 41 AGNs in this study have distance measurements independent of
their redshifts --- NGC\,3227, NGC\,3783, NGC\,4051, NGC\,4151, and
NGC\,4593 --- while for the remaining 36 we estimate the distance from
the redshift of the AGN.  The distance measurements for the five
aforementioned objects generally come from an average of the distance
moduli for galaxies within the same group and were generated as part
of a study of the ``local'' velocity anomaly \citep{tully08}; they
were retrieved from the Extragalactic Distance Database
\citep{tully09}.  They are calibrated to the same zeropoint as the
\hst\ Key Project \citep{freedman01}, which found $H_0 =
72$\,km\,s$^{-1}$\,Mpc$^{-1}$, and which we have adopted throughout
this work.

In general, we find that the uncertainties in the distances are
underestimated for these individual sources in the Extragalactic
Distance Database.  For each of these five objects, we next give a
brief discussion of the available distance measurements and their
apparent quality, as well as the distances we adopt.

\paragraph{NGC\,3227}  
There are seven galaxies in the same group as NGC\,3227, with distance
measurements ranging from $18-34$\,Mpc.  Fortuitously, however,
NGC\,3227 is currently interacting with the early-type galaxy
NGC\,3226.  The distance to NGC\,3226 from the surface brightness
fluctuations (SBF) method is $23.5 \pm 2.4$\,Mpc \citep{tonry01},
which is 10\% less than the group-averaged distance estimate to
NGC\,3227 of $26.4 \pm 1.6$\,Mpc.  We adopt the distance measurement
of NGC\,3226 as the distance to NGC\,3227.

\paragraph{NGC\,3783}
The three galaxies in the group to which NGC\,3783 belongs have
measured distances ranging from 20 to 28\,Mpc, leading to a
group-averaged distance estimate of $25.1 \pm 2.9$\,Mpc for NGC\,3783.
Based on its redshift of 0.00973, however, NGC\,3783 is estimated
to lie at a distance of 41\,Mpc.  This is a difference of nearly 50\%
in distance that translates into a factor of almost 3 difference in
predicted luminosity.  With a recessional velocity of
2917\,km\,s$^{-1}$, NGC\,3783 would generally be expected to have
peculiar velocities affecting its perceived recessional velocity at
only the $\sim10$\% level ($\sim 300$\,km\,s$^{-1}$, e.g.,
\citealt{masters06,bahcall96}), a severe underestimate given the 50\%
discrepancy between the group-averaged distance estimate and the
estimate based on redshift.  We adopt the group-averaged distance of
25.1\,Mpc, with an uncertainty of 20\% (5.0\,Mpc), for NGC\,3783.

\paragraph{NGC\,4051}
NGC\,4051 is one of 64 galaxies identified as belonging to the same
group.  The Tully-Fisher distance to NGC\,4051 is quoted as 12.2\,Mpc,
but it does not appear to have been corrected for the contribution of
the AGN to the total galaxy luminosity.  The AGN contribution would
appear to make the galaxy brighter, and it would therefore seem to be
nearer than it actually is.  The group-averaged distance of $17.1 \pm
0.8$\,Mpc includes individual galaxy distances ranging from
$10-30$\,Mpc.  Within the group of 64 galaxies, there is one galaxy
with a Cepheid distance and eight early-type galaxies with distances
from SBFs.  The distances for these nine
galaxies, which are expected to be more accurate on an individual
basis than the distances to the other 55 galaxies in the group, span a
smaller range of 10--21\,Mpc, with a median value of 14.3\,Mpc that
is fairly consistent with the average distance found for the full
group of 64 galaxies.  Based on our limited information regarding the
location of NGC\,4051 within its group, we adopt the group-averaged
distance of $17.1$\,Mpc for NGC\,4051 based on all the galaxies in the
same group, with an uncertainty of 3.4\,Mpc (20\%).

\paragraph{NGC\,4151}
There are only four galaxies contributing to the group-averaged
distance for NGC\,4151, and their individual distances are estimated
to range from 3.9\,Mpc to 34.0\,Mpc based on the \citet{tully77} line
width-luminosity correlation, with a final distance for NGC\,4151
quoted as $11.2 \pm 1.1$\,Mpc. The object with the smallest estimated
distance of 3.9\,Mpc is NGC\,4151 itself, but the total galaxy
luminosity does not appear to have been corrected for the enormous
contribution from the AGN.  It therefore appears that the distance of
3.9\,Mpc is a gross underestimate caused by neglecting the AGN
contribution to the total galaxy luminosity, causing the galaxy to
appear brighter (and therefore nearer) than it actually is.  We have
recalculated the group-averaged distance while excluding the likely
erroneous distance of 3.9\,Mpc and adopt a distance of $16.6
\pm 3.3$\,Mpc for NGC\,4151.

\paragraph{NGC\,4593}
Only two galaxies contribute to the group-averaged distance for
NGC\,4593.  They have individual distance measurements
of 33\,Mpc and 43\,Mpc.  We adopt the distance estimated by averaging
their distance moduli and an uncertainty of 20\% ($37.3
\pm 7.5$\,Mpc) for NGC\,4593.  This is consistent with the distance
of $\sim 39$\,Mpc expected from the redshift of NGC\,4593 and assuming
that NGC\,4593 has zero peculiar velocity.  The 20\% uncertainty in
the distance that we have assumed for NGC\,4593 (and, indeed, several
of the other distance estimates above) may be an underestimate of the
true discrepancy between the actual distance to the source and our
estimate of the distance.

For the other 36 AGNs in the sample, we have no choice at this time
but to estimate their distances from their measured redshifts.
Because of this, peculiar velocities can introduce a large uncertainty
into these distance estimates.  To further complicate the issue,
peculiar velocities are highly direction-dependent (e.g., the ``Finger
of God'' effect) and will be randomly oriented relative to our line of
sight for the AGN host galaxies in this sample.  Our lack of
additional distance information for the vast majority of the AGNs in
the reverberation sample leads us to conservatively estimate that
peculiar velocities affect the galaxy recession velocities at an
average level of $\sim 500$\,km\,s$^{-1}$, or $\sim 17$\% for
$z=0.01$.  We caution that this may still be a significant
underestimate of the accuracy of our assumed distances for some
individual galaxies, as it would be in the above case of NGC\,3783 if
we had no information beyond the galaxy's redshift.

Clearly, there is a desperate need for accurate distance measurements
to the AGN host galaxies in the reverberation-mapping sample.  The
Tully-Fisher method has been shown to be accurate to $\sim 20$\% for
individual galaxies and can reach spiral galaxies out to $z \approx
0.1$, but will require extra care for these galaxies to ensure removal
of the AGN contribution to the total galaxy luminosity.  Furthermore,
there are a handful of galaxies in the sample that are within reach of
the $\sim 30$\,Mpc limit for Cepheid observations with \hst.  We
have an approved Cycle 20 program to obtain a Cepheid-based distance
measurement to the face-on spiral galaxy NGC\,6814 (GO-12961, PI
Bentz).  NGC\,4151, in particular, is another galaxy with a very large
distance uncertainty that would benefit from, and be within the reach
of, an \hst\ Cepheid program.

Finally, we note that we do not attempt to correct for internal
reddening from the AGN host galaxy.  Previous studies
\citep{bentz09b,denney10} have shown that such corrections, for the
few objects where they are possible, are fairly small relative to the
large distance uncertainties we have described above.  In the case of
the reddened AGN NGC\,3227, for example, the reddening curve derived
by \citet{crenshaw01} gives an extinction of 0.26\,dex in luminosity
at 5100\,\AA.

Table~\ref{tab:rl} lists the 5100\,\AA\ luminosities we have
determined for each of the datasets in our sample using the distances
discussed above.  We also give the corresponding broad H$\beta$ time
delays, which we take to be the average radius of the
H$\beta$-emitting BLR.

\section{The Radius--Luminosity Relationship}

Based on previous work (\citealt{kaspi00,kaspi05,bentz06a,bentz09b}),
we expect the form of the \rl\ relationship to be a power law.  We
parametrize the \rl\ relationship here as
\begin{equation}
\log (R_{\rm BLR}/1~{\rm lt\mbox{-}day}) = K + \alpha \log(\lambda L_{\lambda} / 10^{44}~{\rm ergs\,s^{-1}}).
\end{equation}
\noindent To determine the best fit to the \rl\ relationship, we
employed the {\sc linmix\_err} algorithm \citep{kelly07}, which takes
a Bayesian approach to linear regression with measurement errors in
both coordinates and a component of intrinsic, random scatter.
\citet{kelly07} carried out extensive tests of the consistency between
{\sc linmix\_err} and the commonly used algorithms FITEXY
\citep{press92} and BCES \citep{akritas96}, finding that the
best fits determined by all of the algorithms were generally
consistent, but that even in cases of large scatter or poorly
constrained measurements, {\sc linmix\_err} always derived a fit
that was consistent with the known parent population from which
the measurements were sampled.  We fit the \rl\ relationship
with FITEXY and BCES and found that all the algorithms provided 
consistent results, as we expected.  We report the best-fit
parameters determined by the {\sc linmix\_err} algorithm in
Table~\ref{tab:rlfits}.

The issue of dealing with multiple measurements for a single object
when fitting the \rl\ relationship is not straightforward.  On the one
hand, if an individual AGN moves along its own \rl\ relationship that
is parallel to the \rl\ relationship investigated here, then there is
no real difference between multiple measurements of a single object
versus measurements of many different objects.  In this case, all
measurements should be given equal weight in the regression analysis,
regardless of which AGN they ``belong'' to.  If, on the other hand,
individual AGNs have individual \rl\ relationships that are oriented
at some other slope relative to the \rl\ relationship for the
population, then each AGN should only be allowed to contribute a
single measurement to the regression analysis.  This issue has been
examined for the AGN NGC\,5548 by \citet{peterson02} and
\citet{bentz07}, and both studies found that the slope of the optical
\rl\ relationship for NGC\,5548 alone is steeper than the global
relationship. Preliminary work by Kilerci Eser et al.\ (in
preparation) is also finding that the slope of the optical
\rl\ relationship is steeper for an individual object with multiple
measurements and that the scatter introduced into the global
relationship appears to be modest, but these results are necessarily
based on only the small number of targets with multiple time lag
measurements.  We have also, therefore, investigated the effect of
randomly choosing only a single measurement to represent each object
in the sample, and the best-fit \rl\ relationship is consistent within
the uncertainties with the best-fit relationship derived with all the
measurements for every AGN included.

Figure~\ref{fig:rl} displays the current version of the H$\beta$
\rl\ relationship. In the top-left panel, all of the individual
measurements are plotted.  The new measurements that have been added
to the \rl\ relationship in this work are shown as open circles and
preferentially populate the low-luminosity end of the
relationship.

\subsection{Notes on Individual Objects}

\paragraph{Mrk\,142}  
The narrow-line Seyfert 1 galaxy Mrk\,142 appears to be a significant
outlier in the \rl\ relationship.  With an AGN luminosity of
$3.5\times10^{43}$\,ergs\,s$^{-1}$, it has a predicted H$\beta$ time
lag of $\sim 20$\,days, compared to its observed time lag of $\sim
3$\,days.  Inspection of the continuum and H$\beta$ light curves
presented by \citet{bentz09c} shows a lack of strong features, such
that no apparent lag can be detected by eye.  Furthermore, the
cross-correlation function for Mrk\,142 shows the lowest significance
for the objects in the LAMP sample with reported lag detections.  We
suggest that the reverberation experiment for Mrk\,142 should be
repeated in an effort to detect a more significant time lag.
Repeating the reverberation experiment will also allow for the
confirmation or contradiction of the outlier status of Mrk\,142.
While it is possible that Mrk\,142 is truly an outlier, it is worth
noting that other narrow-line Seyfert 1s in the reverberation sample
(NGC\,4051, NGC\,4253, NGC\,4748) lie extremely close to their
expected locations, and well within the sample scatter.

\paragraph{Arp\,151}
The tidally distorted galaxy Arp\,151 was one of the most variable
objects in the 2008 LAMP campaign, with a well-determined time lag of
$4.0 \pm 0.5$\,days.  However, there appears to be a problem with the
flux calibration for Arp\,151.  The flux calibration for the LAMP
sample of AGNs was determined by comparing the observed [\ion{O}{3}]
$\lambda 5007$\,\AA\ flux for NGC\,5548 to the known [\ion{O}{3}] flux
from many years of spectrophotometric monitoring. Because of the
unstable nature of weather and the need for many observations over a
long period of time, reverberation datasets rely on the narrow
[\ion{O}{3}] emission lines as internal calibration sources for the
many nonphotometric nights on which data is obtained.  The emission
lines do not vary on the timescales probed in a single monitoring
campaign, and they provide the final multiplicative flux calibration
factor in the spectral pipeline.

The [\ion{O}{3}] flux of NGC\,5548 is well-known because of the many
years that this AGN has been monitored and it was the only
well-studied AGN included in the 2008 LAMP campaign.  Looking at the
measured [\ion{O}{3}] fluxes for every night during the 2008 campaign,
there seemed to be only a single night among the 64 nights of the
campaign where the weather at Lick Observatory was steadily
photometric, providing an accurate [\ion{O}{3}] flux measurement for
NGC\,5548.  Unfortunately, it appears that this may not have actually been the
case, at least during the observations of Arp\,151.  The problem
arises from the fact that any reasonable flux for the AGN in the {\it
  HST} image of Arp\,151 is more than the total continuum flux of
AGN+galaxy derived from the ground-based spectroscopy.  

Does this mean that the process of estimating host-galaxy flux from
{\it HST} imaging is flawed?  Probably not.  The clue comes from
reviewing the final spectroscopic flux calibrations for the LAMP AGNs.
The flux in the mean spectrum for each object required an increase by
a multiplicative factor to match the measured [\ion{O}{3}] flux from
the single photometric night.  In the case of Arp\,151 (and only
Ar,151), however, and only Arp\,151, the [\ion{O}{3}] flux measured
from the supposed photometric night of observations is {\it less} than
the mean [\ion{O}{3}] flux from all of the observations.  Reducing the
mean flux by the derived multiplicative factor results in a mean
continuum flux that is too small and thus produces a {\it negative}
AGN flux when the host-galaxy starlight correction is applied.  If we
instead set the multiplicative factor equal to 1, the problem
disappears.  Based on the multiplicative correction factors for all of
the other objects in the LAMP sample, we should expect that this
factor is $\gtrsim 1$ for Arp\,151 .

Indeed, in reviewing the weather logs from the 2008 LAMP campaign and
focusing only on nights logged as possibly photometric and with
[\ion{O}{3}] fluxes that are consistent with each other, we derive a
flux of $0.76\times10^{-13}$\,ergs\,s$^{-1}$\,cm$^{-2}$.  This is
56\% larger than the originally derived [\ion{O}{3}] flux of
$0.49\times10^{-13}$\,ergs\,s$^{-1}$\,cm$^{-2}$.  Furthermore, it is
relatively consistent with the [\ion{O}{3}] flux of
$0.83\times10^{-13}$\,ergs\,s$^{-1}$\,cm$^{-2}$ derived from
potentially photometric nights during the 2011 LAMP \citep{barth11}
monitoring of Arp\,151, given the slightly larger extraction width
adopted during that campaign (10\farcs3 compared to the 9\farcs4
extraction width used in the 2008 campaign) and the overall
uncertainty in the flux calibration for this single object.

We therefore adopt a flux correction factor of 1.56 as derived above,
implying that the mean continuum flux density at rest-frame
5100\,\AA\ for Arp\,151 was $(1.835 \pm 0.079) \times
10^{-15}$\,ergs\,s$^{-1}$\,cm$^{-2}$\,\AA\ during the LAMP 2008
campaign.  We include this corrected measurement in the values
tabulated in Table~\ref{tab:flux}.

\paragraph{PG\,2130+099}
There have been several reverberation experiments targeting
PG\,2130+099, yet there remains some ambiguity as to the accuracy of
the reported lags.  \citet{kaspi00} determined a time lag of $\sim
168$\,days, but a reanalysis by \citet{grier08} found evidence for
aliasing based on seasonal gaps in the light curve.  Analysis of
individual seasonal light curves gave much shorter time lags
(16--44\,days).  Two new reverberation experiments have led to
reported time lags of 23\,days \citep{grier08} and 10\,days
\citep{grier12b}.  The latter experiment is the most recent and has
the best time sampling for this object to date: $\Delta t_{\rm
  med}=0.5$\,days for the continuum and $\Delta t_{\rm med}=1.0$\,days
for the emission line.  With a monitoring baseline of $\sim
120$\,days, however, and a predicted time lag of $\sim 40$\,days based
on the luminosity of PG\,2130+099, in retrospect the experiment is
uncomfortably close to the minimum time baseline recommended for
reverberation experiments \citep{horne04}.  The most obvious feature
in the continuum light curve occurs at a heliocentric Julian day of
HJD$-2450000 \approx 5510$ (\citealt{grier12b} Figure 2) and would be
expected to be echoed in the H$\beta$ light curve at HJD~$\approx
5550$, right where the campaign abruptly ends.  Furthermore,
\citet{grier12c} were able to reconstruct a map of the time-delay
response as a function of velocity across the emission-line profile
using the light curves presented by \citet{grier12b}.  From this
analysis, they determine that the H$\beta$ time lag associated with
this monitoring dataset is likely to be $\sim 31$ days, not 13 days.
It is therefore debatable whether PG\,2130+099 is truly an outlier,
and so we recommend that yet another reverberation experiment be
dedicated to this object.  The tightness of the \rl\ relationship to
date implies that the discovery of a true outlier may well give
important clues about detailed AGN physics deep within the potential
well of the central black hole.

Because of the plausibly erroneous nature of the BLR radius
measurements for Mrk\,142 and PG\,2130+099, we have also carried out
fits to the \rl\ relationship with Mrk\,142 excluded and with an
adopted lag of $31 \pm 4$\,days for PG\,2130+099 (based on the
analysis of \citealt{grier12c}).  With these two changes, the slope of
the relationship is slightly increased, but still consistent within the
uncertainties (see Table~\ref{tab:rlfits}).  We have also considered
the fit with both Mrk\,142 and PG\,2130+099 excluded (see
Table~\ref{tab:rlfits}) and the fit is still consistent within the
uncertainties.

Finally, we have also investigated the effect of correcting for
internal extinction in the one AGN where we observe a large reddening
and we have available an appropriate reddening curve, NGC\,3227.  As
previously mentioned, work by \citet{crenshaw01} leads to an
extinction correction of 0.26\,dex at 5100\,\AA\ for NGC\,3227.  We
have applied this correction to the luminosity measurement for
NGC\,3227, both as a member of the full sample of 71 measurements, and
with the different treatments of Mrk\,142 and PG\,2130+099 described
above.  The results are given in Table~\ref{tab:rlfits}, and
again, in both of these cases, the changes to the best-fit solution
are minimal.  NGC\,3227 is known to be one of the most heavily
reddened objects in our sample, thus, the internal extinction
correction will be much smaller for the rest of the AGNs.  The top
right panel of Figure~\ref{fig:rl} displays the \rl\ relationship with
Mrk\,142 excluded, an adopted lag of $31 \pm 4$\,days for
PG\,2130+099, and a reddening correction for NGC\,3227.

\section{Discussion} 

The best-fit slope of $\alpha = 0.533^{+0.035}_{-0.033}$ is consistent with
the analyses previously presented by \citet{bentz06a,bentz09b}.  There
appears to be no difference in the relationship at the high-luminosity
and low-luminosity ends, with no evidence for a turnover at low
luminosities.  In fact, the \rl\ relationship appears to be remarkably
consistent over four orders of magnitude in luminosity among these
AGNs.  Furthermore, the relationship is remarkably consistent with the
expectation from simple photoionization arguments.  Specifically, it
was first pointed out by \citet{davidson72} that we can define the
ionization parameter of a BLR cloud as
\begin{equation}
U=\frac{Q(H)}{4\pi R^2cn_e},
\end{equation}
\noindent where $R$ is the distance from the central source, $c$ is
the speed of light, $n_e$ is the electron number density, and
\begin{equation}
Q(H) = \int^{\infty}_{\nu_1} \frac{L_{\nu}}{h\nu} d\nu
\end{equation}
\noindent is the flux of hydrogen ionizing photons emitted by the
central source.  Under the assumptions that the ionization parameters
and particle densities are about the same for all AGNs, one finds that
\begin{equation}
R \propto Q(H)^{1/2},
\end{equation}
\noindent so that the radius at which a particular emission line is
most likely to be emitted is a simple function of the intensity of the
ionizing flux.  Further assuming that the ionizing continuum shape is
not a function of luminosity, such that $L \propto Q(H)$, we expect
\begin{equation}
R \propto L^{1/2}.
\end{equation}  
\noindent The above arguments certainly gloss over many of the finer
details of BLR photoionization physics, but this seems to matter
little as the observed relationship matches this simplistic
expectation quite well.  This particular prediction of photoionization
physics, namely that the size of the BLR should scale with the
luminosity of the central source, was sought in the very early days of
reverberation-mapping experiments, when the first BLR sizes were being
measured \citep{koratkar91b}.  It took another decade, however, for
the BLR measurements to span a sufficiently large dynamic range so
that the relationship was clearly detected, despite the large initial
scatter \citep{kaspi00}.

More recent work on the physical basis for an \rl\ relationship,
spurred on by the initial and continuing successes of the
reverberation-mapping method including near-infrared (IR) photometric
reverberation mapping \citep{suganuma06}, has focused on the role of
dust and the dust sublimation radius in setting the size of the BLR.
The importance of dust was first noted by \citet{netzer93} and has
been analyzed more recently by \citet{goad12}, among others.  The
upshot of many of these models is that the outer edge of the BLR is
bounded by the dust sublimation radius, perhaps coincident with the
inner edge of the dusty torus-like structure of the unified model
\citep{antonucci93}.  Outside the dust sublimation radius, the line
emission from the dusty gas is suppressed by a large factor because
the dust grains absorb many of the incoming ionizing photons as well
as the emitted line photons, effectively creating an outer edge for
the BLR.  A natural consequence of a central ionizing source with a
variable flux is that the dust sublimation radius will respond to
these flux variations.  An increase in ionizing photons will destroy
many dust grains and increase the dust sublimation radius, whereas a
decrease in ionizing flux will allow more grains to condense or
migrate in and decrease the dust sublimation radius.  While the basic
physical motivation for a \rl\ relationship seems to be understood,
there are many details that are currently unknown.  \citet{goad12}
provide a comprehensive overview of the state of photoionization
models and the agreement (or lack, thereof) with observations.

The form of the relationship appears to be fairly well-determined at
this point and has not changed significantly with the updates and
additions included here.  The regression results are also consistent
with the results of a microlensing analysis of the BLR in lensed
quasars, where the magnification amplitude is dependent on the size of
the emission region, an independent method that is not subject to the
same uncertainties involved in reverberation mapping
\citep{guerras12}.  Furthermore, the scatter about the
\rl\ relationship is now quite low.
The {\sc linmix\_err} routine provides an estimate of the scatter
about the relationship of $0.19 \pm 0.02$\,dex (about 56\%).  We plot
the residuals of the measured BLR radii to the estimated BLR radii
derived from the results of the {\sc linmix\_err} routine fit in the
bottom left panel of Figure~\ref{fig:rl}.  The residuals are
approximately normally distributed, as can be seen by comparison to
an overplotted Gaussian function with $\sigma=0.19$\,dex (dotted
line).  The scatter about the relationship is often called the
``intrinsic'' scatter, but in this case it is actually a combination
of the real intrinsic scatter and variance from inaccurate or biased
measurements.  Thus, the intrinsic scatter in the \rl\ relationship is
likely to be less than the $0.19 \pm 0.02$\,dex we find here.
Indeed, if we omit the two most suspicious measurements in the sample,
those of Mrk\,142 and PG\,2130+099 as discussed above, we find that
the scatter drops to 0.13\,dex.  The residuals with the exclusion of
Mrk\,142, the adopted lag of $31 \pm 4$ days for PG\,2130+099, and
the reddening correction for NGC\,3227 are plotted in the bottom-right
panel of Figure~\ref{fig:rl}.  This is significantly lower than the
typical scatter in black hole scaling relationships, such as the
0.4\,dex scatter (or larger) of the \msigma\ relationship
\citep{gultekin09,park12,mcconnell12}.

Such low scatter in the \rl\ relationship and the potential to
decrease it even further with accurate distances and additional
reverberation campaigns seems to lend support to the recent arguments
of \citet{watson11} that the \rl\ relationship could potentially be
used to turn any AGN with a well-determined H$\beta$ time lag into a
standardizable candle for use in cosmological studies.  While there is
still work to be done before the \rl\ relationship can match the
0.05\,dex scatter in the Type Ia supernova Hubble diagram residuals
\citep[e.g.,][]{silverman12,ganeshalingam13}, the main weakness at the
moment is the lack of accurate distance measurements to tie the
current sample of reverberation-mapped AGNs onto the well-established
nearby distance ladder.  However, once this has been rectified, the
radius and flux of any AGN could be measured and compared to the
expected luminosity distance derived from the \rl\ relationship.  In
principle, this is possible out to $z \approx 4$ for near-IR
spectroscopy of the H$\beta$ emission line, far beyond the current
reach of Type Ia supernovae.  The high luminosities of such quasars,
however, mean that the observed H$\beta$ time lags would be on the
order of decades. Because the BLR is ionization stratified, emissions
lines with a higher ionization potential than H$\beta$, such as
\ion{C}{4} or \ion{He}{2}, have rest-frame time delays that are
factors of a few smaller than those of H$\beta$.  If separate
\rl\ relationships for these lines could be defined and calibrated as
accurately as we have achieved for H$\beta$ (see \citealt{kaspi07} for
current progress on the \ion{C}{4} \rl\ relationship), then exploring
cosmology with reverberation experiments would become much more
feasible for objects with $z > 2$, where the expected observed time
delays suffer heavily from time-dilation effects.  The ability to
probe out to such high redshifts would provide observational
constraints on the evolution of the dark energy equation-of-state
parameter, as well as on alternative theories of gravity (King et
al.\, in preparation).

\section{Future Work}

The most pressing deficiency in the H$\beta$ \rl\ relationship is the
current lack of accurate distance measurements to the AGN host
galaxies.  The uncertainty in the distances at present provides the
single largest source of uncertainty in the AGN luminosity
measurements, especially as \citet{tully08} have shown that peculiar
velocities may still be important ($>10$\%) even beyond 50\,Mpc.  We
are working to obtain distances based on the Tully-Fisher method for
the intermediate inclination spirals in our sample with new HI
spectroscopy and near-infrared imaging.  We will also obtain a
Cepheid-based distance for NGC6814 (HST GO-12961, PI Bentz), and will
explore additional distance indicators (such as the globular cluster
luminosity function and the planetary nebula luminosity function) for
the remaining AGN host galaxies in the sample.  New reverberation
campaigns should be dedicated to studying the two most likely outliers
in the \rl\ relationship --- PG\,2130+099 and Mrk\,142 --- to
determine whether they are truly outliers.  Additionally, the
difficulties with the flux calibration of Arp\,151 suggest that it
would be a useful exercise to acquire new spectra of all the objects
in the sample, taken under stable photometric conditions and with a
uniform setup.

There is also much work to be done to determine independent \rl\
relationships for \ion{C}{4} and other emission lines for use in
estimating black hole masses at higher redshifts, where the H$\beta$
emission line has redshifted out of the observed optical bandpass.
All of the derived \rl\ and black hole mass scaling relationships for
all other emission lines in the literature currently rely on the
H$\beta$ \rl\ relationship.  It is therefore critical that we build up
independent \rl\ relationships for these other commonly utilized
emission lines.  While \ion{Mg}{2} and \ion{C}{4} are frequently
employed for black hole mass estimates at $z \gtrsim 0.5$, there are
only a handful of existing \ion{C}{4} reverberation results (see
\citealt{kaspi07}), with the vast majority centered around a very
small range in luminosities, and in the case of \ion{Mg}{2}, only a
single object has a measured reverberation time lag.  Furthermore, the
\ion{C}{4} time lags are generally deduced from {\it International
Ultraviolet Explorer (IUE)} spectra that were obtained every few days,
and so have relatively poor temporal sampling compared to what is
typically achieved for H$\beta$ \citep{peterson04}.  The higher
ionization state of \ion{C}{4} means that we expect its time lag to be
a factor of 2--3 shorter than that of H$\beta$.  Future \ion{C}{4}
reverberation experiments of low- to moderate-luminosity AGNs in the
nearby universe will require daily sampling or better in order to
measure \ion{C}{4} time delays to the same level of significance that
is now typically achieved for H$\beta$.

Finally, although this empirical relationship is well measured and
there seems to be some theoretical understanding behind it, the
details of the photoionization physics, as well as the geometry and
kinematics of the gas, are not well understood at this time.  There is
significant room for improvement in our physical understanding of AGN
BLRs.

\section{Summary}

We have carried out an imaging program with \hst\ to provide starlight
corrections to the luminosities of 9 AGNs with H$\beta$ radius
measurements.  We have fully updated and revised the calibration
sample for the \rl\ relationship, including 20 new H$\beta$ BLR
measurements from recent reverberation-mapping campaigns, and we have
reinvestigated the form of the relationship.  We find a best fit of
$\log (R_{\rm BLR}/1 {\rm lt\mbox{-}day}) = 1.527^{+0.031}_{-0.031} +
0.533^{+0.035}_{-0.033} \log(\lambda L_{\lambda} / 10^{44}\,{\rm
  L}_{\odot})$.  This is consistent with a slope of 0.5 and with
previous work that included starlight corrections to the AGN
luminosity measurements.  After including the additions and updates,
the single largest source of uncertainty comes from the highly
uncertain distances to the AGNs in the sample.  The low scatter in the
relationship ($0.19 \pm\ 0.02$\,dex) and the potential to further
reduce the scatter, with no clear outliers, support the proposed use
of the \rl\ relationship to probe the matter and energy content of the
Universe out to $z \approx 0.6$ with optical measurements of the
H$\beta$ emission line.  Pushing the H$\beta$ observations into the
near-IR would allow the relationship to probe quasars out to $z
\approx 4$, beyond the reach of Type Ia supernovae and into a new
interesting regime for tests of the predictions of different
cosmological models.

\acknowledgements 

This work is based on observations with the NASA/ESA {\it Hubble Space
  Telescope}.  We are grateful for support of this work through grant
\hst\ GO-11662 from the Space Telescope Science Institute, which is
operated by the Association of Universities for Research in Astronomy,
Inc., under NASA contract NAS5-26555.  K.D.D. has received funding
from the People Programme (Marie Curie Actions) of the European
Union's Seventh Framework Programme FP7/2007-2013/ under REA grant
agreement no.\ 300553.  A.J.B. acknowledges support from NSF grant
AST-1108835. B.M.P., C.J.G., G.D.R., and R.W.P. acknowledge support
from NSF grant AST-1008882 to Ohio State University.  A.V.F. is
grateful for the support of NSF grants AST-1108665 and AST-1211916,
the TABASGO Foundation, and the Christopher R. Redlich Fund.  The work
of D.S. was carried out at Jet Propulsion Laboratory, California
Institute of Technology, under a contract with NASA.
J.H.W. acknowledges the support by the National Research Foundation of
Korea (NRF) grant funded by the Korea government (No. 2012-006087).
This research has made use of the NASA/IPAC Extragalactic Database
(NED) which is operated by the Jet Propulsion Laboratory, California
Institute of Technology, under contract with the National Aeronautics
and Space Administration and the SIMBAD database, operated at CDS,
Strasbourg, France.  We dedicate this paper to the memory of our dear
friend and colleague, Weidong Li, whose tireless dedication to the
Katzman Automatic Imaging Telescope (KAIT) significantly contributed
to the success of LAMP; his premature, tragic passing has deeply
saddened us.

\clearpage

\bibliographystyle{apj} 

\clearpage

\begin{figure*}
\caption{\hst\ WFC3 F547M images of the AGN host galaxies, displayed
  with an inverted logarithmic stretch. The black rectangles show the
  geometry and orientation of each ground-based spectroscopic
  monitoring aperture. The size of the region displayed is
  1\arcmin$\times$1\arcmin, except for NGC\,6814 which is displayed in
  a 2\arcmin$\times$2\arcmin\ box.  For all images, north is up and
  east is to the left. }
\label{fig:apertures}
\end{figure*}

\begin{figure*}
\caption{Image (top left), models (middle), and residuals (right) for
  Mrk\,142.  The upper panels display the ``simple'' models and their
  residuals, and the bottom panels display the ``optimal'' models and
  their residuals.  The scale of the black bar in the top-left panel
  is 10\,arcsec and the compass in the bottom corner of the panel
  shows the directions north and east. The galaxy image and models are
  displayed with an inverted logarithmic stretch, and the residuals
  are displayed with an inverted linear stretch centered around zero
  counts.  The bottom-left panel shows the one-dimensional surface
  brightness of the galaxy (data points), the best-fit model (solid
  line), and each of the individual best-fit model components (PSF =
  dotted line, all others = dashed lines), with the ellipticity of the
  galaxy displayed below.}
\label{fig:m142}
\end{figure*}

\begin{figure*}
\caption{Same as Figure~2, but for SBS\,1116+583A.}
\label{fig:s1116}
\end{figure*}

\begin{figure*}
\caption{Same as Figure~2, but for Arp\,151.  The ``O'' shapes in the
images are reflections in the optics from a nearby bright object.}
\label{fig:arp151}
\end{figure*}

\begin{figure*}
\caption{Same as Figure~2, but for Mrk\,1310.}
\label{fig:m1310}
\end{figure*}

\begin{figure*}
\caption{Same as Figure~2, but for Mrk\,202.}
\label{fig:m202}
\end{figure*}

\begin{figure*}
\caption{Same as Figure~2, but for NGC\,4253 (Mrk\,766).}
\label{fig:n4253}
\end{figure*}

\begin{figure*}
\caption{Same as Figure~2, but for NGC\,4748.}
\label{fig:n4748}
\end{figure*}

\begin{figure*}
\caption{Same as Figure~2, but for Mrk\,290. The small jump in surface
brightness at 9\arcsec\ is due to the bright field star in the image.}
\label{fig:m290}
\end{figure*}

\begin{figure*}
\caption{Same as Figure~2, but for NGC\,6814.}
\label{fig:n6814}
\end{figure*}

\begin{figure*}
\centering
\begin{minipage}{0.495\linewidth}
\centering
\plotone{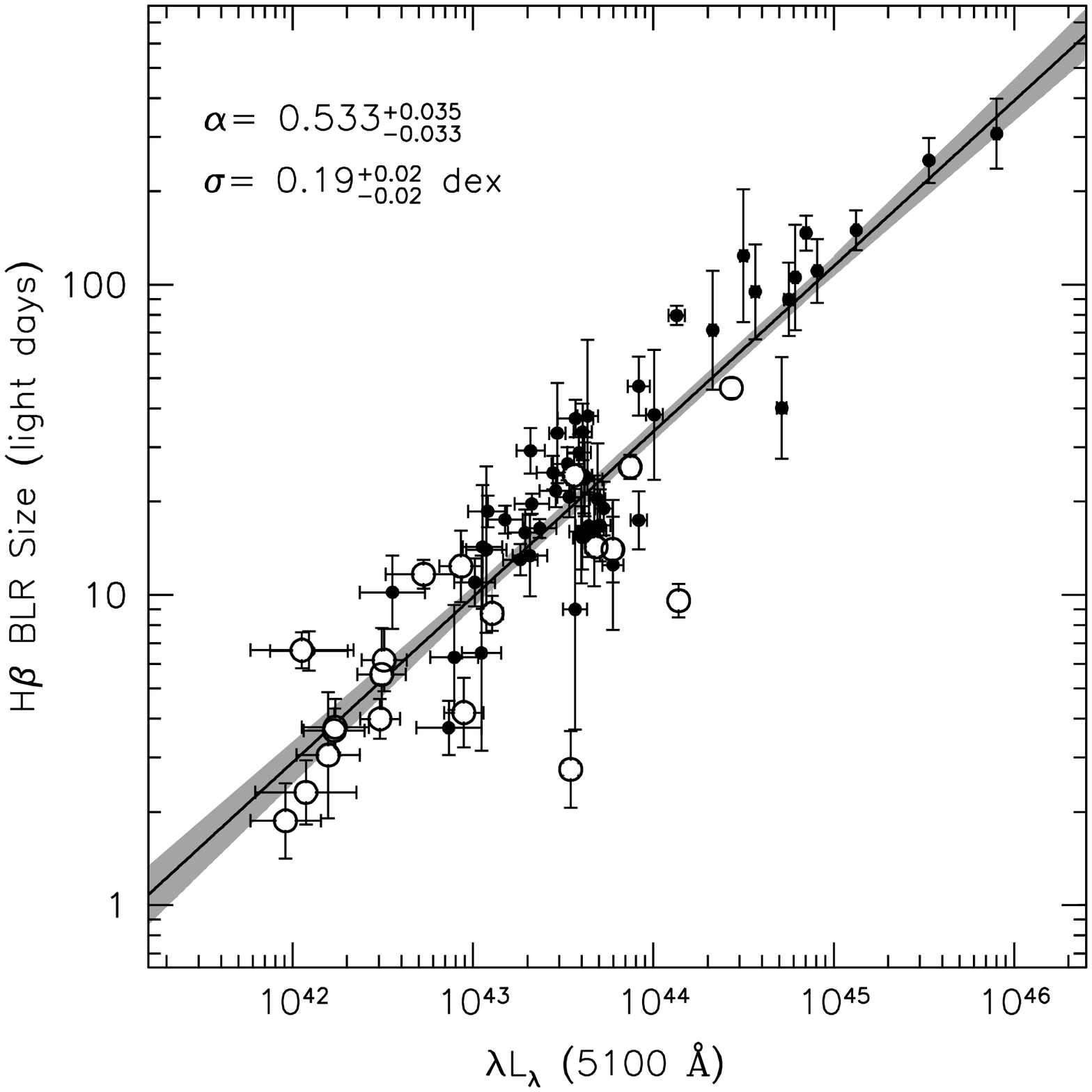} \\
\plotone{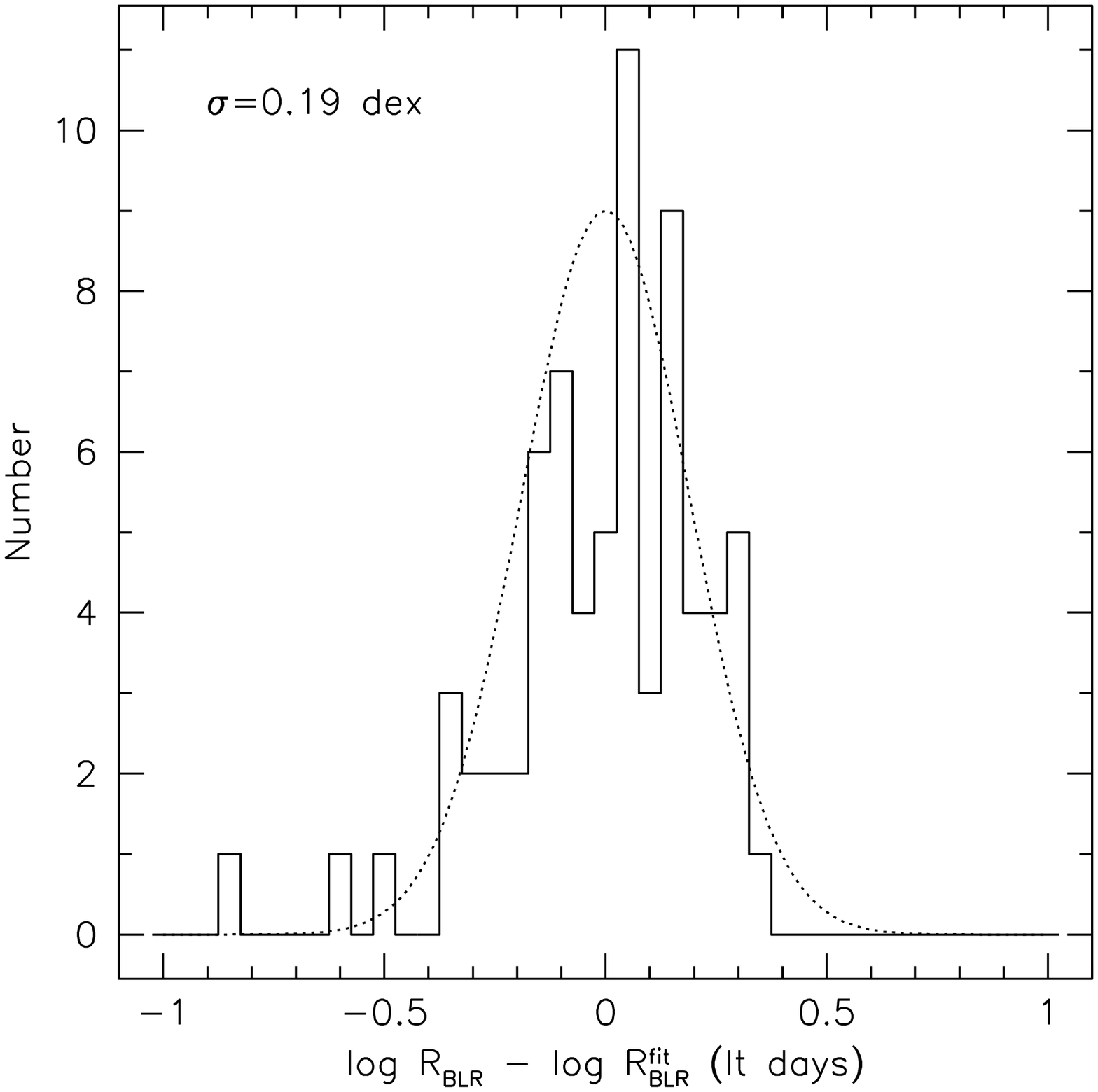}
\end{minipage}
\begin{minipage}{0.495\linewidth}
\plotone{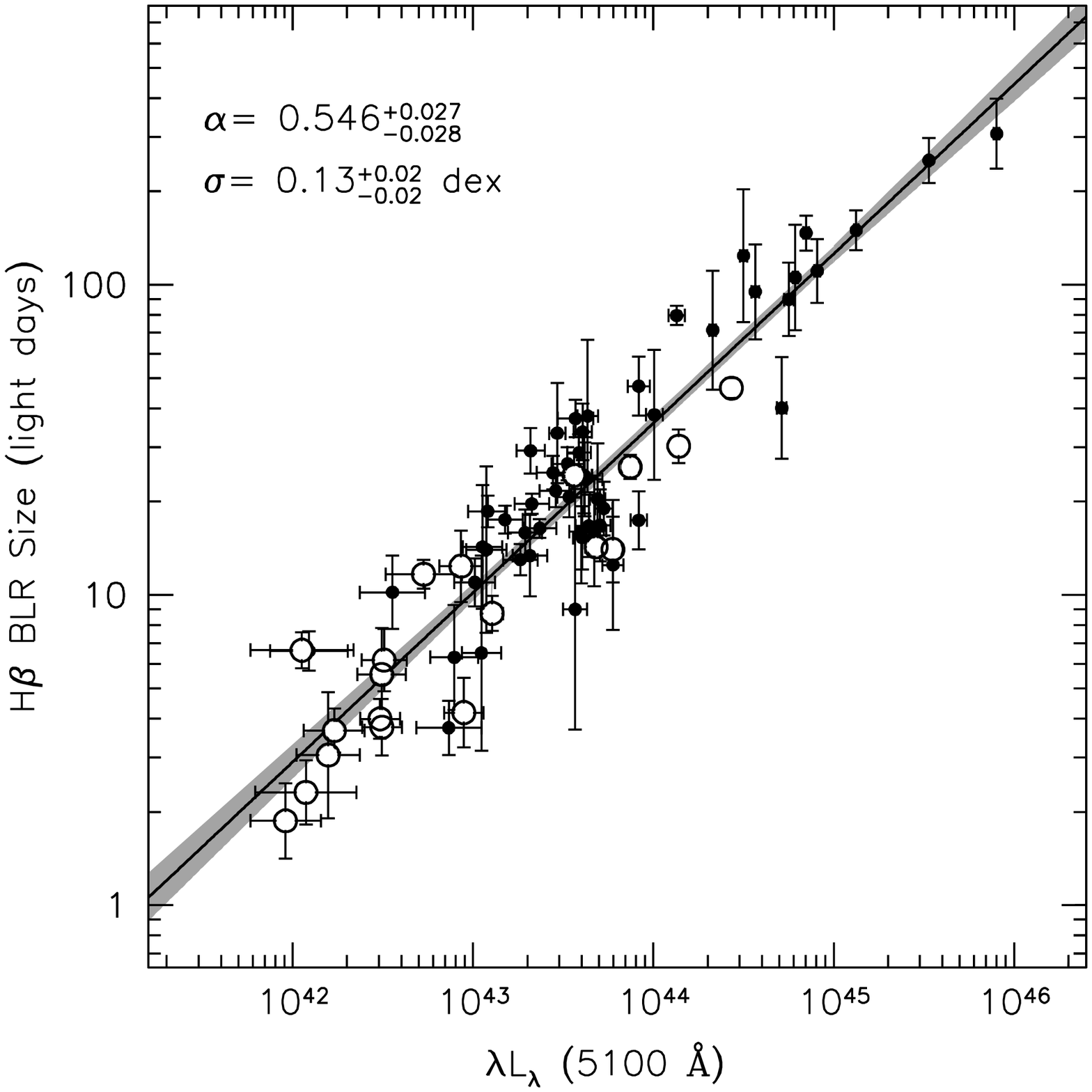} \\
\plotone{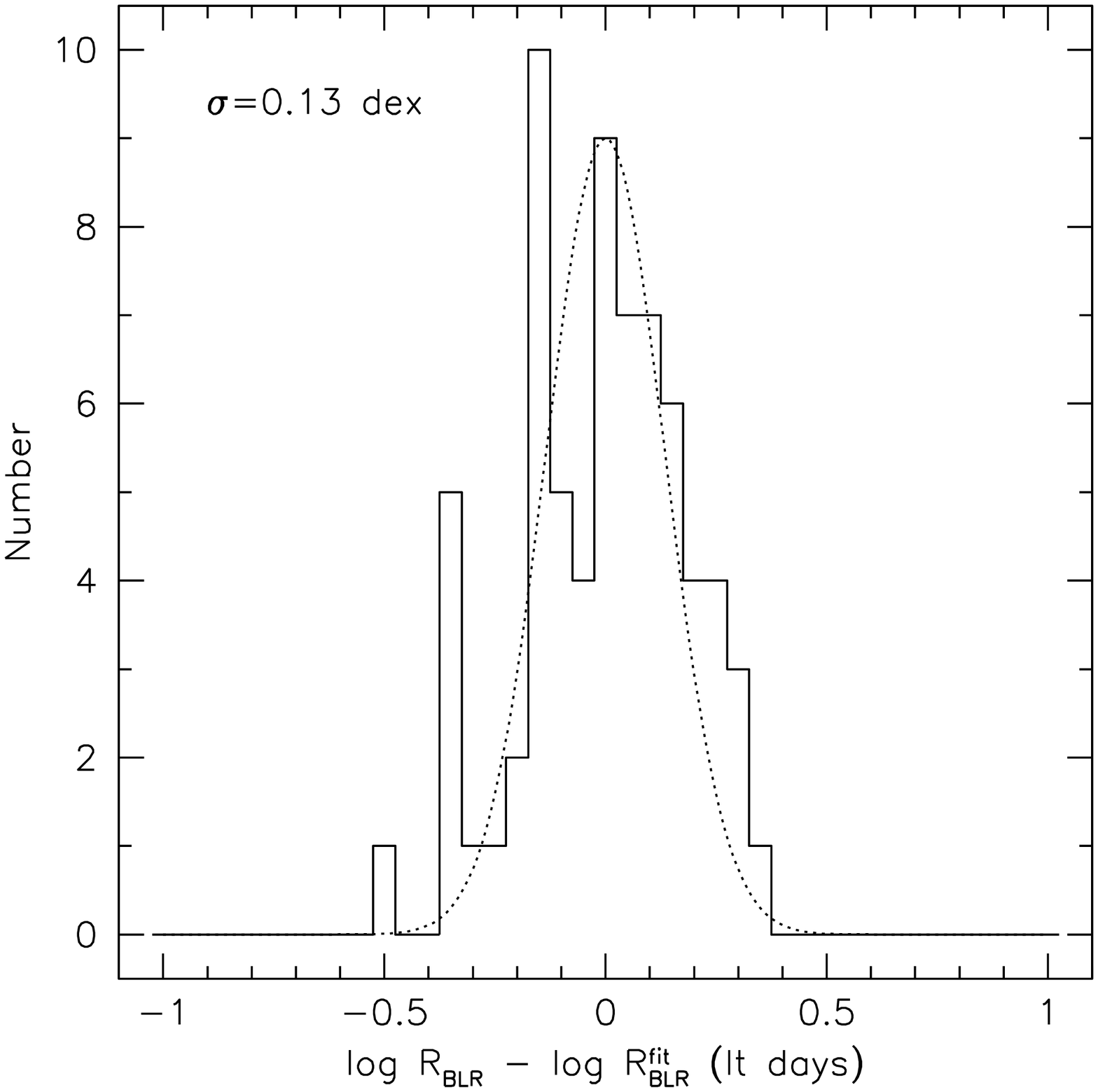}
\end{minipage}
\caption{{\it Top:} H$\beta$ BLR radius versus the
  5100\,\AA\ AGN luminosity.  The solid line is the best fit
  to the data and the grayscale region shows the range allowed by the
  uncertainties on the best fit.  The left panel displays all 71
  datapoints included in this analysis, where the open circles are the
  new measurements that we include for the first time.  The right
  panel shows the fit with Mrk\,142 removed, an adopted lag for
  PG\,2130+099 of $31 \pm 4$\,days, and a reddening correction of
  0.26\,dex for NGC\,3227 (see the text for details).  The slope does
  not change appreciably with these adjustments, but the scatter is
  significantly reduced from 0.19\,dex to 0.13\,dex.  All measurements
  are plotted with their associated uncertainties, but the error bars
  are sometimes smaller than the plot symbols. {\it Bottom:} Residuals
  of the estimated BLR radii compared to the measured BLR radii using
  the best fit to the \rl\ relationship.  The dotted lines are
  Gaussian functions with a width equal to the variance in the scatter
  determined from the best fit, demonstrating the relative normality 
  of the residual distribution.}
\label{fig:rl}
\end{figure*}

\clearpage

\begin{deluxetable}{lllcccl}
\tablecolumns{7}
\tablewidth{0pt}
\tabletypesize{\scriptsize}
\tablecaption{Object List}
\tablehead{
\colhead{Object} &
\colhead{$\alpha_{\rm J2000}$} &
\colhead{$\delta_{\rm J2000}$} &
\colhead{$z$} &
\colhead{$D_{\rm L}$\tablenotemark{a}} &
\colhead{$A_{B}$\tablenotemark{b}} &
\colhead{Alternate} \\
\colhead{} &
\colhead{(hr min sec)} &
\colhead{(\arcdeg\ \arcmin\ \arcsec)} &
\colhead{} &
\colhead{(Mpc)} &
\colhead{(mag)} &
\colhead{Name}}

\startdata
Mrk\,335       & 00\,06\,19.5   & +20\,12\,10       & 0.02579  & 113           & 0.129  & PG\,0003+199 \\
PG\,0026+129   & 00\,29\,13.6   & +13\,16\,03       & 0.14200  & 672           & 0.258  & \\
PG\,0052+251   & 00\,54\,52.1   & +25\,25\,38       & 0.15500  & 740           & 0.172  & \\
Fairall\,9     & 01\,23\,45.8   & $-$58\,48\,21     & 0.04702  & 209           & 0.097  & \\
Mrk\,590       & 02\,14\,33.6   & $-$00\,46\,00     & 0.02639  & 115           & 0.136  & NGC\,863 \\
3C\,120	       & 04\,33\,11.1   & +05\,21\,16       & 0.03301  & 145           & 1.078  & Mrk\,1506 \\
Ark\,120       & 05\,16\,11.4   & $-$00\,08\,59     & 0.03271  & 142           & 0.466  & Mrk\,1095 \\
Mrk\,79        & 07\,42\,32.8   & +49\,48\,35       & 0.02219  & 97            & 0.257  & \\
\enddata

\tablenotetext{a}{Distances were estimated from the redshifts of the
  AGNs except for five cases: NGC\,3227, NGC\,3783, NGC\,4051,
  NGC\,4151, and NGC\,4593.  The distances and uncertainties for the
  last four of these objects are from an average of distance moduli to
  neighboring galaxies in the \citet{tully08} study of the local
  velocity anomaly, with distance uncertainties estimated at 20\%.
  The distance to NGC\,3227 is based on the SBF method \citep{tonry01}
  for NGC\,3226, with which NGC\,3227 is currently
  interacting. Additional details are given in the text.}
                  
\tablenotetext{b}{Values are from the \citet{schlafly11} recalibration
  of the \citet{schlegel98} dust map.}

\end{deluxetable}

\begin{deluxetable}{lccc}
\tablecolumns{4}
\tablewidth{0pt}
\tabletypesize{\scriptsize}
\tablecaption{\hst\ Observation Log}
\tablehead{
\colhead{Object} &
\colhead{Date Observed} &
\colhead{Total Exposure} &
\colhead{Roll Angle} \\
\colhead{} &
\colhead{(yyyy--mm--dd)} &
\colhead{Time (s)} &
\colhead{(\degr E of N)}
}

\startdata
Mrk\,142         & 2010-04-22  & 2450  & $-$8.29  \\
SBS\,1116+583A   & 2010-06-06  & 2510  & $-$28.1  \\
Arp\,151         & 2010-04-09  & 2450  &    29.7  \\
Mrk\,1310        & 2009-12-02  & 2240  &    157.1 \\
Mrk\,202         & 2010-04-14  & 2510  &    25.5  \\
Mrk\,766         & 2010-06-21  & 2270  & $-$22.5  \\
NGC\,4748        & 2010-06-28  & 2250  & $-$24.1  \\
Mrk\,290         & 2010-07-25  & 2520  & $-$22.0  \\
NGC\,6814        & 2010-05-06  & 2240  &    120.7 \\

\label{tab:hstobs}
\enddata                            

\end{deluxetable}


\begin{deluxetable}{lclcccccccl}
\tablecolumns{11}
\tablewidth{0pt}
\tabletypesize{\scriptsize}
\tablecaption{Surface Brightness Decomposition --- Mrk\,142}
\tablehead{
\colhead{Fit} &
\colhead{\#} &
\colhead{PSF+sky} &
\colhead{$\Delta x$ (\arcsec)} &
\colhead{$\Delta y$ (\arcsec)} &
\colhead{$m_{\rm stmag}$\tablenotemark{a}} &
\colhead{...} &
\colhead{Sky (cts)} &
\colhead{$\frac{d{\rm sky}}{dx}$ ($10^{-4}$ cts)} &
\colhead{$\frac{d{\rm sky}}{dy}$ ($10^{-4}$ cts)} &
\colhead{Note} \\
\colhead{} &
\colhead{} &
\colhead{sersic} &
\colhead{$\Delta x$ (\arcsec)} &
\colhead{$\Delta y$ (\arcsec)} &
\colhead{$m_{\rm stmag}$} &
\colhead{$r_{\rm e}$ (\arcsec)} &
\colhead{$n$} &
\colhead{$b/a$} &
\colhead{PA (deg)} &
\colhead{} \\
\colhead{} &
\colhead{} &
\colhead{power} &
\colhead{\nodata} &
\colhead{$r_{\rm in}$ (\arcsec)} &
\colhead{$r_{\rm out}$ (\arcsec)} &
\colhead{$\theta_{\rm rot}$ (deg)} &
\colhead{$\alpha$} &
\colhead{$\theta_{\rm incl}$ (deg)} &
\colhead{$\theta_{\rm sky}$ (deg)} &
\colhead{} \\
\colhead{} &
\colhead{} &
\colhead{fourier} &
\colhead{\nodata} &
\multicolumn{2}{c}{mode: $a_{\rm m}$ , $\phi$ (deg)} &
\multicolumn{2}{c}{mode: $a_{\rm m}$ , $\phi$ (deg)} &
\multicolumn{2}{c}{mode: $a_{\rm m}$ , $\phi$ (deg)} &
\colhead{} \\
\colhead{(1)} &
\colhead{(2)} &
\colhead{(3)} &
\colhead{(4)} &
\colhead{(5)} &
\colhead{(6)} &
\colhead{(7)} &
\colhead{(8)} &
\colhead{(9)} &
\colhead{(10)} &
\colhead{(11)}
}
\startdata
Optimal & 1,2 & PSF+sky   & 0	 & 0	     & 16.10  &	\nodata	  & 33.2   & 6.1	& -7.5  & \\
     & 3   & sersic    & 0.12	 & -0.10     & 18.78  &	0.17	  & [1.0]  & 0.12       & 118.7 & \\
     & 	   & ~~power   & \nodata & 0.22	     & 0.38   & -94.5     & 0.4367 & 61.6       & 2.1   & \\
     & 	   & ~~fourier & \nodata & 1: 0.90   & -70.1  & 3: 0.22	  & 0.5	   & 4:	-0.038  & -6.1  & \\
     & 	   & ~~fourier & \nodata & 5: 0.028  & -1.1   &		  & 	   &	        &       & \\
     & 4   & sersic    & 0.30	 & 0.29	     & 19.44  &	0.22	  & 0.6	   & 0.34       & 33.2  & \\
     & 	   & ~~fourier & \nodata & 1: -0.31  & 99.5   & 3: 0.193  & -2.9   & 4:	0.2468  & 34.8  & \\
     & 	   & ~~fourier & \nodata & 5: 0.12   & 28.3   &		  & 	   &	        &       & \\
\enddata  
\tablecomments{Values in square brackets were held fixed during the surface brightness model fitting.}
\tablenotetext{a}{The STmag magnitude system is based on the absolute physical flux per unit wavelength.}
\end{deluxetable}

\begin{deluxetable}{lclcccccccl}
\tablecolumns{11}
\tablewidth{0pt}
\tabletypesize{\scriptsize}
\tablecaption{Surface Brightness Decomposition --- SBS\,1116+583A}
\tablehead{
\colhead{Fit} &
\colhead{\#} &
\colhead{PSF+sky} &
\colhead{$\Delta x$ (\arcsec)} &
\colhead{$\Delta y$ (\arcsec)} &
\colhead{$m_{\rm stmag}$\tablenotemark{a}} &
\colhead{...} &
\colhead{Sky (cts)} &
\colhead{$\frac{d{\rm sky}}{dx}$ ($10^{-4}$ cts)} &
\colhead{$\frac{d{\rm sky}}{dy}$ ($10^{-4}$ cts)} &
\colhead{Note} \\
\colhead{} &
\colhead{} &
\colhead{sersic} &
\colhead{$\Delta x$ (\arcsec)} &
\colhead{$\Delta y$ (\arcsec)} &
\colhead{$m_{\rm stmag}$} &
\colhead{$r_{\rm e}$ (\arcsec)} &
\colhead{$n$} &
\colhead{$b/a$} &
\colhead{PA (deg)} &
\colhead{} \\
\colhead{} &
\colhead{} &
\colhead{power} &
\colhead{\nodata} &
\colhead{$r_{\rm in}$ (\arcsec)} &
\colhead{$r_{\rm out}$ (\arcsec)} &
\colhead{$\theta_{\rm rot}$ (deg)} &
\colhead{$\alpha$} &
\colhead{$\theta_{\rm incl}$ (deg)} &
\colhead{$\theta_{\rm sky}$ (deg)} &
\colhead{} \\
\colhead{} &
\colhead{} &
\colhead{fourier} &
\colhead{\nodata} &
\multicolumn{2}{c}{mode: $a_{\rm m}$ , $\phi$ (deg)} &
\multicolumn{2}{c}{mode: $a_{\rm m}$ , $\phi$ (deg)} &
\multicolumn{2}{c}{mode: $a_{\rm m}$ , $\phi$ (deg)} &
\colhead{} \\
\colhead{(1)} &
\colhead{(2)} &
\colhead{(3)} &
\colhead{(4)} &
\colhead{(5)} &
\colhead{(6)} &
\colhead{(7)} &
\colhead{(8)} &
\colhead{(9)} &
\colhead{(10)} &
\colhead{(11)}
}
\startdata
Optimal  & 1,2 & PSF+sky     & 0.0     & 0.0       & 18.02      & \nodata   & 40.57      & -7.24     & -26.04     &   \\
      & 3   & sersic      & -0.04   & 0.06      & 17.31      & 1.01      & 1.1        & 0.86      & 39.4       &   \\
      &     &  ~~fourier  & \nodata & 1: 0.08   & 1: 103.04  & 3: 0.01   & 3: -89.96  & 4: -0.02  & 4: 5.84    &   \\
      &     &  ~~fourier  & \nodata & 5: -0.00  & 5: -31.95  &           &            &           &            &   \\
      & 4   & sersic      & -0.26   & 1.99      & 16.81      & 5.31      & 0.3        & 0.83      & -69.6      &   \\
      &     &  ~~fourier  & \nodata & 1: 0.24   & 1: 18.19   & 3: -0.03  & 3: 0.67    & 4: 0.01   & 4: 9.84    &   \\
      &     &  ~~fourier  & \nodata & 5: 0.01   & 5: -23.10  &           &            &           &            &   \\
      & 5   & sersic      & -0.28   & 0.03      & 17.37      & 7.92      & [1.0]      & 0.34      & 42.5       &   \\
\enddata  
\tablecomments{Values in square brackets were held fixed during the surface brightness model fitting.}
\tablenotetext{a}{The STmag magnitude system is based on the absolute physical flux per unit wavelength.}
\end{deluxetable}

\begin{deluxetable}{lclcccccccl}
\tablecolumns{11}
\tablewidth{0pt}
\tabletypesize{\scriptsize}
\tablecaption{Surface Brightness Decomposition --- Arp\,151}
\tablehead{
\colhead{Fit} &
\colhead{\#} &
\colhead{PSF+sky} &
\colhead{$\Delta x$ (\arcsec)} &
\colhead{$\Delta y$ (\arcsec)} &
\colhead{$m_{\rm stmag}$\tablenotemark{a}} &
\colhead{...} &
\colhead{Sky (cts)} &
\colhead{$\frac{d{\rm sky}}{dx}$ ($10^{-4}$ cts)} &
\colhead{$\frac{d{\rm sky}}{dy}$ ($10^{-4}$ cts)} &
\colhead{Note} \\
\colhead{} &
\colhead{} &
\colhead{sersic} &
\colhead{$\Delta x$ (\arcsec)} &
\colhead{$\Delta y$ (\arcsec)} &
\colhead{$m_{\rm stmag}$} &
\colhead{$r_{\rm e}$ (\arcsec)} &
\colhead{$n$} &
\colhead{$b/a$} &
\colhead{PA (deg)} &
\colhead{} \\
\colhead{} &
\colhead{} &
\colhead{fourier} &
\colhead{\nodata} &
\multicolumn{2}{c}{mode: $a_{\rm m}$ , $\phi$ (deg)} &
\multicolumn{2}{c}{mode: $a_{\rm m}$ , $\phi$ (deg)} &
\multicolumn{2}{c}{mode: $a_{\rm m}$ , $\phi$ (deg)} &
\colhead{} \\
\colhead{(1)} &
\colhead{(2)} &
\colhead{(3)} &
\colhead{(4)} &
\colhead{(5)} &
\colhead{(6)} &
\colhead{(7)} &
\colhead{(8)} &
\colhead{(9)} &
\colhead{(10)} &
\colhead{(11)}
}

\startdata 
Optimal	& 1,2 &	PSF+sky	  &  0	    &	0	  &  17.25   &	\nodata	 &  24.3  &  6.0      &	-3.2	&	\\
	& 3   &	PSF 	  &  0.00   &	0.04	  &  17.55   &		 &  	  &	      &		&	\\
	& 4   &	sersic	  &  -0.10  &	-0.04	  &  16.86   &	1.40	 &  4.7	  &  0.47     &	-40.7	&	\\
	&     &	~~fourier & \nodata &  1: -0.51	  &  57.4    & 3: 0.16	 &  27.7  & 4: 0.065  &	-11.3	&	\\
	&     &	~~fourier & \nodata &  5: -0.024  &  15.4    &		 &  	  &	      &		&	\\
	& 5   &	sersic	  &  0.09   &	-0.13	  &  15.99   &	2.87	 &  2.5	  &  0.52     &	-60.7	&	\\
	&     &	~~fourier & \nodata &  1: 0.10	  &  50.1    & 3: 0.10	 &  -34.7 & 4: 0.025  &	-10.0	&	\\
	&     &	~~fourier & \nodata &  5: -0.018  &  31.3    &		 &  	  &	      &		&	\\
\enddata  
\tablecomments{Values in square brackets were held fixed during the surface brightness model fitting.}
\tablenotetext{a}{The STmag magnitude system is based on the absolute physical flux per unit wavelength.}
\end{deluxetable}

\begin{deluxetable}{lclcccccccl}
\tablecolumns{11}
\tablewidth{0pt}
\tabletypesize{\scriptsize}
\tablecaption{Surface Brightness Decomposition --- Mrk\,1310}
\tablehead{
\colhead{Fit} &
\colhead{\#} &
\colhead{PSF+sky} &
\colhead{$\Delta x$ (\arcsec)} &
\colhead{$\Delta y$ (\arcsec)} &
\colhead{$m_{\rm stmag}$\tablenotemark{a}} &
\colhead{...} &
\colhead{Sky (cts)} &
\colhead{$\frac{d{\rm sky}}{dx}$ ($10^{-4}$ cts)} &
\colhead{$\frac{d{\rm sky}}{dy}$ ($10^{-4}$ cts)} &
\colhead{Note} \\
\colhead{} &
\colhead{} &
\colhead{sersic} &
\colhead{$\Delta x$ (\arcsec)} &
\colhead{$\Delta y$ (\arcsec)} &
\colhead{$m_{\rm stmag}$} &
\colhead{$r_{\rm e}$ (\arcsec)} &
\colhead{$n$} &
\colhead{$b/a$} &
\colhead{PA (deg)} &
\colhead{} \\
\colhead{} &
\colhead{} &
\colhead{sersic3} &
\colhead{$\Delta x$ (\arcsec)} &
\colhead{$\Delta y$ (\arcsec)} &
\colhead{$\Sigma_{\rm b}$} &
\colhead{$r_{\rm b}$ (\arcsec)} &
\colhead{$n$} &
\colhead{$b/a$} &
\colhead{PA (deg)} &
\colhead{} \\
\colhead{} &
\colhead{} &
\colhead{radial} &
\colhead{$\Delta x$ (\arcsec)} &
\colhead{$\Delta y$ (\arcsec)} &
\colhead{\nodata} &
\colhead{$r_{\rm break}$ (\arcsec)} &
\colhead{$\Delta r_{\rm soft}$ (\arcsec)} &
\colhead{$b/a$} &
\colhead{PA (deg)} &
\colhead{} \\
\colhead{} &
\colhead{} &
\colhead{power} &
\colhead{\nodata} &
\colhead{$r_{\rm in}$ (\arcsec)} &
\colhead{$r_{\rm out}$ (\arcsec)} &
\colhead{$\theta_{\rm rot}$ (deg)} &
\colhead{$\alpha$} &
\colhead{$\theta_{\rm incl}$ (deg)} &
\colhead{$\theta_{\rm sky}$ (deg)} &
\colhead{} \\
\colhead{} &
\colhead{} &
\colhead{fourier} &
\colhead{\nodata} &
\multicolumn{2}{c}{mode: $a_{\rm m}$ , $\phi$ (deg)} &
\multicolumn{2}{c}{mode: $a_{\rm m}$ , $\phi$ (deg)} &
\multicolumn{2}{c}{mode: $a_{\rm m}$ , $\phi$ (deg)} &
\colhead{} \\
\colhead{(1)} &
\colhead{(2)} &
\colhead{(3)} &
\colhead{(4)} &
\colhead{(5)} &
\colhead{(6)} &
\colhead{(7)} &
\colhead{(8)} &
\colhead{(9)} &
\colhead{(10)} &
\colhead{(11)}
}

\startdata
Optimal	& 1,2 &	PSF+sky	  & 0	    &	0	 & 17.42     &	\nodata	  & [73.5] &	5.9	&  -3.2	  & \\
	& 3   &	PSF 	  & -0.03   &	0.02	 & 19.45     &		  & 	   &		&  	  & \\
	& 4   &	PSF 	  & -0.15   &	0.00	 & 21.70     &		  & 	   &		&  	  & \\
	& 5   &	sersic	  & -0.02   &	0.00	 & 16.15     & 3.29	  & 4.8	   &	0.80	&  163.0  & \\
	&     &	~~fourier & \nodata & 1: -0.075	 & 70.9	     & 3: -0.0086 & -42.9  & 4:	0.0092	&  6.5	  & \\
	&     &	~~fourier & \nodata & 5: -0.0050 & 21.7	     & 6: 0.0093  & -26.8  &		&  	  & \\
	& 6   &	sersic	  & -0.053  &	0.014	 & 15.34     & 4.84	  & [1.0]  &	0.86	&  -99.9  & \\
	&     &	~~power	  & \nodata &	1.73	 & 3.57	     & 365.9	  & 0.050  &	32.1	&  72.9	  & \\
\enddata  

\tablecomments{Values in square brackets were held fixed during the surface brightness model fitting.}
\tablenotetext{a}{The STmag magnitude system is based on the absolute physical flux per unit wavelength.}
\end{deluxetable}

\begin{deluxetable}{lclcccccccl}
\tablecolumns{11}
\tablewidth{0pt}
\tabletypesize{\scriptsize}
\tablecaption{Surface Brightness Decomposition --- Mrk\,202}
\tablehead{
\colhead{Fit} &
\colhead{\#} &
\colhead{PSF+sky} &
\colhead{$\Delta x$ (\arcsec)} &
\colhead{$\Delta y$ (\arcsec)} &
\colhead{$m_{\rm stmag}$\tablenotemark{a}} &
\colhead{...} &
\colhead{Sky (cts)} &
\colhead{$\frac{d{\rm sky}}{dx}$ ($10^{-4}$ cts)} &
\colhead{$\frac{d{\rm sky}}{dy}$ ($10^{-4}$ cts)} &
\colhead{Note} \\
\colhead{} &
\colhead{} &
\colhead{sersic} &
\colhead{$\Delta x$ (\arcsec)} &
\colhead{$\Delta y$ (\arcsec)} &
\colhead{$m_{\rm stmag}$} &
\colhead{$r_{\rm e}$ (\arcsec)} &
\colhead{$n$} &
\colhead{$b/a$} &
\colhead{PA (deg)} &
\colhead{} \\
\colhead{} &
\colhead{} &
\colhead{sersic3} &
\colhead{$\Delta x$ (\arcsec)} &
\colhead{$\Delta y$ (\arcsec)} &
\colhead{$\Sigma_{\rm b}$} &
\colhead{$r_{\rm b}$ (\arcsec)} &
\colhead{$n$} &
\colhead{$b/a$} &
\colhead{PA (deg)} &
\colhead{} \\
\colhead{} &
\colhead{} &
\colhead{radial} &
\colhead{$\Delta x$ (\arcsec)} &
\colhead{$\Delta y$ (\arcsec)} &
\colhead{\nodata} &
\colhead{$r_{\rm break}$ (\arcsec)} &
\colhead{$\Delta r_{\rm soft}$ (\arcsec)} &
\colhead{$b/a$} &
\colhead{PA (deg)} &
\colhead{} \\
\colhead{} &
\colhead{} &
\colhead{power} &
\colhead{\nodata} &
\colhead{$r_{\rm in}$ (\arcsec)} &
\colhead{$r_{\rm out}$ (\arcsec)} &
\colhead{$\theta_{\rm rot}$ (deg)} &
\colhead{$\alpha$} &
\colhead{$\theta_{\rm incl}$ (deg)} &
\colhead{$\theta_{\rm sky}$ (deg)} &
\colhead{} \\
\colhead{} &
\colhead{} &
\colhead{fourier} &
\colhead{\nodata} &
\multicolumn{2}{c}{mode: $a_{\rm m}$ , $\phi$ (deg)} &
\multicolumn{2}{c}{mode: $a_{\rm m}$ , $\phi$ (deg)} &
\multicolumn{2}{c}{mode: $a_{\rm m}$ , $\phi$ (deg)} &
\colhead{} \\
\colhead{(1)} &
\colhead{(2)} &
\colhead{(3)} &
\colhead{(4)} &
\colhead{(5)} &
\colhead{(6)} &
\colhead{(7)} &
\colhead{(8)} &
\colhead{(9)} &
\colhead{(10)} &
\colhead{(11)}
}

\startdata
Optimal  &	1,2 & PSF+sky	& 0	  & 0	      &	18.13  & \nodata    & 34.4  & 5.8	 & -7.8	 &	\\
      &	3   & sersic	& 0.01    & 0.02      &	16.47  & 0.48	    & 2.2   & 0.79	 & -80.3 &	\\
      &	    & ~~fourier	& \nodata & 1: 0.11   &	74.3   & 3: 0.019   & -6.2  & 4: 0.0054	 & 14.1	 &	\\
      &	    & ~~fourier	& \nodata & 5: 0.0074 &	0.9    & 6: -0.0064 & 6.5   &		 & 	 &	\\
      &	4   & sersic3	& 1.25	  & 2.74      &	9.34   & 0.61	    & 0.8   & 0.61	 & -28.1 &	\\
      && ~~radial,inner & -0.60   & -46.35    &        & 1.65       & 0.80  & 0.66       & -99.1 &      \\
      &	    & ~~fourier & \nodata & 1: 0.51   &	159.9  & 3: -0.17   & 40.5  & 4: 0.13	 & 27.3	 &	\\
      &	    & ~~fourier & \nodata & 5: 0.074  &	-14.3  & 6: -0.041  & -16.5 &		 & 	 &	\\
\enddata  

\tablecomments{Values in square brackets were held fixed during the surface brightness model fitting.}
\tablenotetext{a}{The STmag magnitude system is based on the absolute physical flux per unit wavelength.}
\end{deluxetable}

\begin{deluxetable}{lclcccccccl}
\tablecolumns{11}
\tablewidth{0pt}
\tabletypesize{\scriptsize}
\tablecaption{Surface Brightness Decomposition --- NGC\,4253}
\tablehead{
\colhead{Fit} &
\colhead{\#} &
\colhead{PSF+sky} &
\colhead{$\Delta x$ (\arcsec)} &
\colhead{$\Delta y$ (\arcsec)} &
\colhead{$m_{\rm stmag}$\tablenotemark{a}} &
\colhead{\nodata} &
\colhead{Sky (cts)} &
\colhead{$\frac{d{\rm sky}}{dx}$ ($10^{-4}$ cts)} &
\colhead{$\frac{d{\rm sky}}{dy}$ ($10^{-4}$ cts)} &
\colhead{Note} \\
\colhead{} &
\colhead{} &
\colhead{sersic} &
\colhead{$\Delta x$ (\arcsec)} &
\colhead{$\Delta y$ (\arcsec)} &
\colhead{$m_{\rm stmag}$} &
\colhead{$r_{\rm e}$ (\arcsec)} &
\colhead{$n$} &
\colhead{$b/a$} &
\colhead{PA (deg)} &
\colhead{} \\
\colhead{} &
\colhead{} &
\colhead{sersic3} &
\colhead{$\Delta x$ (\arcsec)} &
\colhead{$\Delta y$ (\arcsec)} &
\colhead{$\Sigma_{\rm b}$} &
\colhead{$r_{\rm b}$ (\arcsec)} &
\colhead{$n$} &
\colhead{$b/a$} &
\colhead{PA (deg)} &
\colhead{} \\
\colhead{} &
\colhead{} &
\colhead{power} &
\colhead{\nodata} &
\colhead{$r_{\rm in}$ (\arcsec)} &
\colhead{$r_{\rm out}$ (\arcsec)} &
\colhead{$\theta_{\rm rot}$ (deg)} &
\colhead{$\alpha$} &
\colhead{$\theta_{\rm incl}$ (deg)} &
\colhead{$\theta_{\rm sky}$ (deg)} &
\colhead{} \\
\colhead{} &
\colhead{} &
\colhead{radial} &
\colhead{$\Delta x$ (\arcsec)} &
\colhead{$\Delta y$ (\arcsec)} &
\colhead{\nodata} &
\colhead{$r_{\rm break}$ (\arcsec)} &
\colhead{$\Delta r_{\rm soft}$ (\arcsec)} &
\colhead{$b/a$} &
\colhead{PA (deg)} &
\colhead{} \\
\colhead{} &
\colhead{} &
\colhead{fourier} &
\colhead{\nodata} &
\multicolumn{2}{c}{mode: $a_{\rm m}$ , $\phi$ (deg)} &
\multicolumn{2}{c}{mode: $a_{\rm m}$ , $\phi$ (deg)} &
\multicolumn{2}{c}{mode: $a_{\rm m}$ , $\phi$ (deg)} &
\colhead{} \\
\colhead{(1)} &
\colhead{(2)} &
\colhead{(3)} &
\colhead{(4)} &
\colhead{(5)} &
\colhead{(6)} &
\colhead{(7)} &
\colhead{(8)} &
\colhead{(9)} &
\colhead{(10)} &
\colhead{(11)}
}
\startdata
Optimal &	1,2 & PSF+sky	& 0	  & 0	     &	15.69	& \nodata  & 28.5  & 2.8      &	-5.4	&	\\
     &	3   & sersic	& -0.12   & 0.01    &	17.01	& 0.22	   & 3.9   & 0.77     &	-4.9	&	\\
     &	    & ~~fourier	& \nodata & 1: -0.25 &	-36.1	& 3: 0.25  & 8.2   & 4:	0.091 &	-30.1	&	\\
     &	4   & sersic	& -3.73	  & -3.76    &	16.82	& 3.71	   & 0.3   & 0.16     &	-44.7	&	\\
     &	    & ~~fourier	& \nodata & 1: 0.34  &	-139.3	& 3: -0.43 & 3.9   & 4:	0.076 &	43.2	&	\\
     &	    & ~~fourier	& \nodata & 5: 0.073 &	8.4	& 6: 0.13  & 4.6   &	      &		&	\\
     &	5   & sersic	& -0.04   & 0.33     &	15.38	& 6.47	   & [1.0] & 0.48     &	30.4	&	\\
     &	    & ~~power	& \nodata & [0.00]   &	12.75	& 199.9	   & -1.27 & 58.4     &	50.0	&	\\
\enddata  

\tablecomments{Values in square brackets were held fixed during the surface brightness model fitting.}
\tablenotetext{a}{The STmag magnitude system is based on the absolute physical flux per unit wavelength.}
\end{deluxetable}

\begin{deluxetable}{lclcccccccl}
\tablecolumns{11}
\tablewidth{0pt}
\tabletypesize{\scriptsize}
\tablecaption{Surface Brightness Decomposition --- NGC\,4748}
\tablehead{
\colhead{Fit} &
\colhead{\#} &
\colhead{PSF+sky} &
\colhead{$\Delta x$ (\arcsec)} &
\colhead{$\Delta y$ (\arcsec)} &
\colhead{$m_{\rm stmag}$\tablenotemark{a}} &
\colhead{\nodata} &
\colhead{Sky (cts)} &
\colhead{$\frac{d{\rm sky}}{dx}$ ($10^{-4}$ cts)} &
\colhead{$\frac{d{\rm sky}}{dy}$ ($10^{-4}$ cts)} &
\colhead{Note} \\
\colhead{} &
\colhead{} &
\colhead{sersic} &
\colhead{$\Delta x$ (\arcsec)} &
\colhead{$\Delta y$ (\arcsec)} &
\colhead{$m_{\rm stmag}$} &
\colhead{$r_{\rm e}$ (\arcsec)} &
\colhead{$n$} &
\colhead{$b/a$} &
\colhead{PA (deg)} &
\colhead{} \\
\colhead{} &
\colhead{} &
\colhead{sersic3} &
\colhead{$\Delta x$ (\arcsec)} &
\colhead{$\Delta y$ (\arcsec)} &
\colhead{$\Sigma_{\rm b}$} &
\colhead{$r_{\rm b}$ (\arcsec)} &
\colhead{$n$} &
\colhead{$b/a$} &
\colhead{PA (deg)} &
\colhead{} \\
\colhead{} &
\colhead{} &
\colhead{power} &
\colhead{\nodata} &
\colhead{$r_{\rm in}$ (\arcsec)} &
\colhead{$r_{\rm out}$ (\arcsec)} &
\colhead{$\theta_{\rm rot}$ (deg)} &
\colhead{$\alpha$} &
\colhead{$\theta_{\rm incl}$ (deg)} &
\colhead{$\theta_{\rm sky}$ (deg)} &
\colhead{} \\
\colhead{} &
\colhead{} &
\colhead{radial} &
\colhead{$\Delta x$ (\arcsec)} &
\colhead{$\Delta y$ (\arcsec)} &
\colhead{\nodata} &
\colhead{$r_{\rm break}$ (\arcsec)} &
\colhead{$\Delta r_{\rm soft}$ (\arcsec)} &
\colhead{$b/a$} &
\colhead{PA (deg)} &
\colhead{} \\
\colhead{} &
\colhead{} &
\colhead{fourier} &
\colhead{\nodata} &
\multicolumn{2}{c}{mode: $a_{\rm m}$ , $\phi$ (deg)} &
\multicolumn{2}{c}{mode: $a_{\rm m}$ , $\phi$ (deg)} &
\multicolumn{2}{c}{mode: $a_{\rm m}$ , $\phi$ (deg)} &
\colhead{} \\
\colhead{(1)} &
\colhead{(2)} &
\colhead{(3)} &
\colhead{(4)} &
\colhead{(5)} &
\colhead{(6)} &
\colhead{(7)} &
\colhead{(8)} &
\colhead{(9)} &
\colhead{(10)} &
\colhead{(11)}
}
\startdata
Optimal	& 1,2	& PSF+sky    &	0	& 0	    & 16.58   &	\nodata	   & [38.5]  &	0.0	 & -0.1	  &	\\
	& 3	& sersic     &	-0.02 	& 0.01      & 16.17   &	0.19	   & 4.8     &	0.83	 & -4.7	  &	\\
	& 	& ~~fourier  & \nodata	& 1: -0.011 & -115.2  & 3: 0.11	   & 45.3    & 4: -0.024 & 2.9	  &	\\
	& 	& ~~fourier  & \nodata  & 5: 0.017  & 1.4     & 6: -0.008  & 7.3     &		 & 	  &	\\
	& 4	& sersic3    &	-0.06	& 0.43	    & 20.96   &	[0.80]	   & [1.0]   &	0.61	 & -87.1  &	\\
	&& ~~radial,inner    &	0.23	& -46.05    & \nodata &	2.62	   & 2.90    &	0.42	 & -83.6  &	\\
	& 	& ~~fourier  &	\nodata	& 1: 0.55   & 43.1    & 3: -0.14   & -17.0   & 4: 0.063	 & 30.4	  &	\\
	& 	& ~~fourier  &	\nodata	& 5: 0.038  & -25.4   & 6: 0.059   & -14.8   &		 & 	  &	\\
\enddata  

\tablecomments{Values in square brackets were held fixed during the surface brightness model fitting.}
\tablenotetext{a}{The STmag magnitude system is based on the absolute physical flux per unit wavelength.}
\end{deluxetable}

\begin{deluxetable}{lclcccccccl}
\tablecolumns{11}
\tablewidth{0pt}
\tabletypesize{\scriptsize}
\tablecaption{Surface Brightness Decomposition --- Mrk\,290}
\tablehead{
\colhead{Fit} &
\colhead{\#} &
\colhead{PSF+sky} &
\colhead{$\Delta x$ (\arcsec)} &
\colhead{$\Delta y$ (\arcsec)} &
\colhead{$m_{\rm stmag}$\tablenotemark{a}} &
\colhead{...} &
\colhead{Sky (cts)} &
\colhead{$\frac{d{\rm sky}}{dx}$ ($10^{-4}$ cts)} &
\colhead{$\frac{d{\rm sky}}{dy}$ ($10^{-4}$ cts)} &
\colhead{Note} \\
\colhead{} &
\colhead{} &
\colhead{sersic} &
\colhead{$\Delta x$ (\arcsec)} &
\colhead{$\Delta y$ (\arcsec)} &
\colhead{$m_{\rm stmag}$} &
\colhead{$r_{\rm e}$ (\arcsec)} &
\colhead{$n$} &
\colhead{$b/a$} &
\colhead{PA (deg)} &
\colhead{} \\
\colhead{} &
\colhead{} &
\colhead{fourier} &
\colhead{\nodata} &
\multicolumn{2}{c}{mode: $a_{\rm m}$ , $\phi$ (deg)} &
\multicolumn{2}{c}{mode: $a_{\rm m}$ , $\phi$ (deg)} &
\multicolumn{2}{c}{mode: $a_{\rm m}$ , $\phi$ (deg)} &
\colhead{} \\
\colhead{(1)} &
\colhead{(2)} &
\colhead{(3)} &
\colhead{(4)} &
\colhead{(5)} &
\colhead{(6)} &
\colhead{(7)} &
\colhead{(8)} &
\colhead{(9)} &
\colhead{(10)} &
\colhead{(11)}
}

\startdata 
Optimal	& 1,2	& PSF+sky   &	0	& 0	    & 15.54  & \nodata	 & 37.2	 & 7.3	     &	-9.0	& 	\\
	& 3	& PSF 	    &	-0.16	& -0.00     & 19.39  &		 & 	 & 	     &		&	\\
	& 4	& PSF 	    &	0.02	& 0.03	    & 16.76  &		 & 	 & 	     &		&	\\
	& 5	& PSF 	    &	-0.14	& -0.11	    & 19.77  &		 & 	 & 	     &		&	\\
	& 6	& PSF 	    &	-0.25	& 8.38	    & 17.79  &		 & 	 & 	     &		& star	\\
	& 7	& sersic    &	0.04	& -0.15	    & 16.48  &	1.25	 & 2.3	 & 0.85	     &	-78.2	&	\\
	& 	& ~~fourier &   \nodata & 1: -0.21  & 90.3   & 3: 0.013	 & -25.6 & 4: 0.029  &	28.6	&	\\
	& 	& ~~fourier &	\nodata & 5: 0.0093 & -5.9   &		 & 	 &	     &		&	\\
\enddata  

\tablecomments{Values in square brackets were held fixed during the surface brightness model fitting.}
\tablenotetext{a}{The STmag magnitude system is based on the absolute physical flux per unit wavelength.}
\end{deluxetable}

\begin{deluxetable}{lclcccccccl}
\tablecolumns{11}
\tablewidth{0pt}
\tabletypesize{\scriptsize}
\tablecaption{Surface Brightness Decomposition --- NGC\,6814}
\tablehead{
\colhead{Fit} &
\colhead{\#} &
\colhead{PSF+sky} &
\colhead{$\Delta x$ (\arcsec)} &
\colhead{$\Delta y$ (\arcsec)} &
\colhead{$m_{\rm stmag}$\tablenotemark{a}} &
\colhead{...} &
\colhead{Sky (cts)} &
\colhead{$\frac{d{\rm sky}}{dx}$ ($10^{-4}$ cts)} &
\colhead{$\frac{d{\rm sky}}{dy}$ ($10^{-4}$ cts)} &
\colhead{Note} \\
\colhead{} &
\colhead{} &
\colhead{sersic} &
\colhead{$\Delta x$ (\arcsec)} &
\colhead{$\Delta y$ (\arcsec)} &
\colhead{$m_{\rm stmag}$} &
\colhead{$r_{\rm e}$ (\arcsec)} &
\colhead{$n$} &
\colhead{$b/a$} &
\colhead{PA (deg)} &
\colhead{} \\
\colhead{} &
\colhead{} &
\colhead{power} &
\colhead{\nodata} &
\colhead{$r_{\rm in}$ (\arcsec)} &
\colhead{$r_{\rm out}$ (\arcsec)} &
\colhead{$\theta_{\rm rot}$ (deg)} &
\colhead{$\alpha$} &
\colhead{$\theta_{\rm incl}$ (deg)} &
\colhead{$\theta_{\rm sky}$ (deg)} &
\colhead{} \\
\colhead{} &
\colhead{} &
\colhead{fourier} &
\colhead{\nodata} &
\multicolumn{2}{c}{mode: $a_{\rm m}$ , $\phi$ (deg)} &
\multicolumn{2}{c}{mode: $a_{\rm m}$ , $\phi$ (deg)} &
\multicolumn{2}{c}{mode: $a_{\rm m}$ , $\phi$ (deg)} &
\colhead{} \\
\colhead{(1)} &
\colhead{(2)} &
\colhead{(3)} &
\colhead{(4)} &
\colhead{(5)} &
\colhead{(6)} &
\colhead{(7)} &
\colhead{(8)} &
\colhead{(9)} &
\colhead{(10)} &
\colhead{(11)}
}
\startdata
Optimal	& 1,2 &	PSF+sky	  &  0	     &	0	& 16.52	  &	\nodata	&  49.4	  &	14.7	&  19.8	   &	\\
	& 3   &	PSF 	  &  0.01    &	0.06	& 17.99	  &		&  	  &		&  	   &	\\
	& 4   &	PSF 	  &  -0.07   &	0.03	& 18.90	  &		&  	  &		&  	   &	\\
	& 5   &	sersic	  &  -0.17   &	0.01	& 15.87	  &	0.90	&  1.5	  &	0.71	&  172.6   &	\\
	&     &	~~fourier &  \nodata & 1: 0.40	& -74.3	  & 3:	0.090	&  40.3	  & 4:	0.049	&  15.7	   &	\\
	& 6   &	sersic	  &  -0.98   &	0.10	& 14.34	  &	4.10	&  1.2	  &	0.83	&  67.7	   &	\\
	&     &	~~fourier &  \nodata & 1: 0.21	& -2.8	  & 3:	0.018	&  1.2	  & 4:	0.026	&  -20.0   &	\\
	& 7   &	sersic	  &  -7.57   &	-0.10	& 12.80	  &	28.42	&  [1.0]  &	0.66	&  -151.3  &	\\
\enddata  
\tablecomments{Values in square brackets were held fixed during the surface brightness model fitting.}
\tablenotetext{a}{The STmag magnitude system is based on the absolute physical flux per unit wavelength.}
\end{deluxetable}


\begin{deluxetable}{lccrclcccc}
\tablecolumns{11}
\tablewidth{0pt}
\tabletypesize{\scriptsize}
\tablecaption{Ground-Based Monitoring Apertures and Measured Flux Densities}
\tablehead{
\colhead{Object} &
\colhead{Ref.\tablenotemark{a}} &
\colhead{$f_{\rm obs}$[5100\,\AA($1+z$)]} &
\multicolumn{3}{c}{Aperture} &
\colhead{PA} &
\colhead{photflam} &
\colhead{$f_{5100}/f_{\rm HST}$} &
\colhead{$f_{\rm gal}$[5100\,\AA($1+z$)]} \\
\colhead{} &
\colhead{} &
\colhead{($10^{-15} f_{\lambda}$)} &
\multicolumn{3}{c}{( \arcsec\ $\times$ \arcsec )} &
\colhead{(\arcdeg)} &
\colhead{$10^{-19}$\,ergs\,cm$^{-2}$\,\AA$^{-1}$\,e$^{-1}$} &
\colhead{} &
\colhead{($10^{-15} f_{\lambda}$)} }

\startdata

Mrk\,335       & 1  &	$7.683	\pm 0.151$  & 5.0  & $\times$ &	7.6	& 90.0	&  $5.838  $ &  0.847  &  $1.559  \pm 0.078$	\\
               & 1  &	$8.809	\pm 0.176$  & 5.0  & $\times$ &	7.6	& 90.0	&  $5.838  $ &  0.847  &  $1.559  \pm 0.078$	\\
               & 2 &	$7.490	\pm 0.268$  & 5.0  & $\times$ &	12.0	& 0.0	&  $5.838  $ &  0.847  &  $1.646  \pm 0.082$	\\
PG\,0026+129   & 3  &	$2.690	\pm 0.060$  & 10.0 & $\times$ &	13.0	& 42.0	&  $76.912 $ &  0.995  &  $0.379  \pm 0.019$	\\
PG\,0052+251   & 3  &	$2.070	\pm 0.065$  & 10.0 & $\times$ &	13.0	& 153.4	&  $5.839  $ &  0.976  &  $0.682  \pm 0.034$	\\
Fairall\,9     & 4  &	$5.950	\pm 0.085$  & 4.0  & $\times$ &	9.0	& 0.0	&  $5.791  $ &  0.888  &  $2.997  \pm 0.150$	\\
Mrk\,590       & 1  &	$7.895	\pm 0.170$  & 5.0  & $\times$ &	7.6	& 90.0	&  $5.796  $ &  0.848  &  $3.965  \pm 0.198$	\\
               & 1  &	$5.331	\pm 0.124$  & 5.0  & $\times$ &	7.6	& 90.0	&  $5.796  $ &  0.848  &  $3.965  \pm 0.198$	\\
\enddata

\tablecomments{Flux densities are tabulated at rest-frame 5100\,\AA\
  in $f_{\lambda}$ units of ergs\,s$^{-1}$\,cm$^{-2}$\,\AA$^{-1}$.
  Position angle is defined as degrees east of north.  {\it Photflam}
  is the inverse sensitivity of the {\it HST} detector, listed here in
  units of $10^{-19}$\,ergs\,cm$^{-2}$\,\AA$^{-1}$\,electron$^{-1}$. }

\tablenotetext{a}{The extraction width of this aperture was previously
reported incorrectly.  The correct extraction width is one pixel
smaller and is as reported here.}

\tablenotetext{b}{The position angle was erroneously reported as
PA~=~90\arcdeg\ for these monitoring campaigns previously.  The actual
position angle is as listed here, PA~=~0\arcdeg.}

\tablerefs{1. \citet{peterson98a},
           2. \citet{grier12b},
	   3. \citet{kaspi00},
	   4. \citet{santoslleo97},
	   5. \citet{denney09b},
           6. \citet{bentz09c},
           7. \citet{stirpe94},				
           8. \citet{bentz06b},				    
           9. \citet{denney06},				    
          10. \citet{winge96},				    
          11. \citet{santoslleo01},			    
          12. \citet{peterson02} and references therein,    
          13. \citet{bentz07},				    
          14. \citet{dietrich98},
	  15. \citet{dietrich12},
          16. \citet{peterson13}.}

\end{deluxetable}


\begin{deluxetable}{lrcc}
\tablecolumns{4}
\tablewidth{0pt}
\tablecaption{Rest-frame H$\beta$ Time Lags and 5100\,\AA\ Luminosities}
\tablehead{
\colhead{Object} &
\colhead{H$\beta$ Time Lag} &
\colhead{$f_{\rm AGN}$ [5100\,\AA\,($1+z$)]} &
\colhead{$\lambda L_{\lambda,{\rm AGN}}$ (5100\,\AA)}\\
\colhead{} &
\colhead{(days)} &
\colhead{($10^{-15}$\,ergs\,s$^{-1}$\,cm$^{-2}$\,\AA$^{-1}$)} &
\colhead{($10^{44}$\,ergs\,s$^{-1}$)}
}

\startdata

Mrk\,335    	&   $16.8  ^{+4.8 }_{-4.2  }$   &   $6.12 \pm  0.19$   &   $43.70  \pm 0.06	$ \\                
 	      	&   $12.5  ^{+6.6 }_{-5.5  }$   &   $7.25 \pm  0.21$   &   $43.78  \pm 0.05	$ \\                
	     	&   $14.3  ^{+0.7 }_{-0.7  }$   &   $5.84 \pm  0.29$   &   $43.68  \pm 0.06	$ \\        
PG\,0026+129    &   $111.0 ^{+24.1}_{-28.3 }$   &   $2.31 \pm  0.07$   &   $44.91  \pm 0.02	$ \\                
PG\,0052+251    &   $89.8  ^{+24.5}_{-24.1 }$   &   $1.39 \pm  0.08$   &   $44.75  \pm 0.03	$ \\                
Fairall\,9      &   $17.4  ^{+3.2 }_{-4.3  }$   &   $2.95 \pm  0.23$   &   $43.92  \pm 0.05	$ \\                
Mrk\,590        &   $20.7  ^{+3.5 }_{-2.7  }$   &   $3.93 \pm  0.33$   &   $43.53  \pm 0.07	$ \\                
	      	&   $14.0  ^{+8.5 }_{-8.8  }$   &   $1.37 \pm  0.31$   &   $43.07  \pm 0.11	$ \\
\enddata													      
				    					  					      
\tablecomments{References for H$\beta$ time lags are in the same order
as the references in Table~\ref{tab:flux}.}

\end{deluxetable}


\begin{deluxetable}{lcccc}
\tablecolumns{5}
\renewcommand{\tabcolsep}{0.7cm}
\renewcommand{\arraystretch}{1.4}
\tablewidth{0pt}
\tablecaption{H$\beta$ \rl\ Fits}
\tablehead{
\colhead{Note} &
\colhead{N} &
\colhead{K} &
\colhead{$\alpha$} &
\colhead{$\sigma^2$} 
}
\startdata 






All            & 71   & $1.527^{+0.031}_{-0.031}$ & $0.533^{+0.035}_{-0.033}$ & $0.036^{+0.009}_{-0.008}$ \\   
Clean          & 70   & $1.555^{+0.024}_{-0.024}$ & $0.542^{+0.027}_{-0.026}$ & $0.018^{+0.006}_{-0.004}$ \\
Clean2         & 69   & $1.560^{+0.024}_{-0.024}$ & $0.546^{+0.027}_{-0.027}$ & $0.018^{+0.006}_{-0.004}$ \\   
ExtCorr        & 71   & $1.526^{+0.030}_{-0.031}$ & $0.537^{+0.034}_{-0.035}$ & $0.037^{+0.010}_{-0.008}$ \\   
Clean+ExtCorr  & 70   & $1.554^{+0.024}_{-0.024}$ & $0.546^{+0.027}_{-0.028}$ & $0.018^{+0.006}_{-0.004}$ \\
Clean2+ExtCorr & 69   & $1.559^{+0.024}_{-0.024}$ & $0.549^{+0.028}_{-0.027}$ & $0.018^{+0.006}_{-0.005}$ \\
\enddata

\tablecomments{The scatter is reported as $\sigma^2$, the variance of
  the measurements around the best-fit relationship.  {\it All:} All
  individual measurements are included in the fit.  {\it Clean:} The
  potentially erroneous time-lag measurement for Mrk\,142 is excluded
  from the fit, and the potentially biased time lag for PG\,2130+099
  is replaced with $\tau = 31 \pm\ 4$\,days.  {\it Clean2}: The lags
  for both Mrk\,142 and PG\,2130+099 are excluded. {\it ExtCorr:} The
  luminosity of NGC\,3227 is corrected for 0.26\,dex of internal
  extinction.}

\end{deluxetable}


\end{document}